\documentclass{emulateapj}
\usepackage{apjfonts}

\newlength{\colwidth}
\newcommand{\cm}{{\rm cm}}
\newcommand{\s}{{\rm s}}
\newcommand{\kms}{{\rm km}\,{\rm s}^{-1}}
\newcommand{\K}{{\rm K}}
\newcommand{\pc}{{\rm pc}}
\newcommand{\kpc}{{\rm kpc}}
\newcommand{\mpc}{{\rm Mpc}}
\newcommand{\Msun}{{{\rm M}_\odot}}

\newcommand{\HI}{\ion{H}{1}} 
\newcommand{\HII}{\ion{H}{2}} 
\newcommand{\Htwo}{{H$_2$}}

\shorttitle{Star formation thresholds and galaxy edges}
\shortauthors{Schaye}

\begin{document}

\submitted{Accepted for publication in the Astrophysical Journal}

\title{Star formation thresholds and galaxy edges: why and where} 
\author{Joop~Schaye}
\affil{School of Natural Sciences, Institute for Advanced
Study, Einstein Drive, Princeton NJ 08540, schaye@ias.edu}

\begin{abstract}
We study global star formation thresholds in the outer parts of
galaxies by investigating the stability of disk galaxies embedded in
dark halos. The disks are self-gravitating, contain metals and dust,
and are exposed to UV radiation. We find that the critical surface
density for the existence of a cold interstellar phase depends only
weakly on the parameters of the model and coincides with the empirically
derived surface density threshold for star formation. Furthermore,
it is shown that the drop in the thermal velocity dispersion
associated with the transition from the warm to the cold gas phase
triggers gravitational instability on a wide range of scales. The
presence of strong turbulence 
does not undermine this conclusion if the disk is
self-gravitating. Models based on the hypothesis that the onset of
thermal instability determines the star formation threshold in the
outer parts of galaxies can reproduce many observations, including the
threshold radii, column densities, and the sizes of stellar disks as a
function of disk scale length and mass. Finally, prescriptions are
given for implementing star formation thresholds in (semi-)analytic
models and three-dimensional hydrodynamical simulations of galaxy
formation.  
\end{abstract}


\section{Introduction}
\label{sec:intro}
Observations of the distribution of H$\alpha$ emission in disk
galaxies show that the (azimuthally averaged) star formation rate
(SFR) drops abruptly at a few disk scale lengths (e.g., Kennicutt
1989, hereafter K89; Martin \& Kennicutt 2001, hereafter MK01). The
fact that the gas disk typically extends far beyond this
radius, suggests that the radial truncation of the
SFR is due to a star formation threshold. If the
truncation radius does not decrease with time and if stars remain at
fixed radii after their formation, then one would also expect a sharp
cutoff in the stellar surface density. There is evidence that
stellar disks in galaxies are finite: the surface brightness
distribution of spiral galaxies is observed to cut off beyond a few
disk scale lengths (van der Kruit 1979; van der Kruit \& Searle
1981). There are some indications that the two disk edges do coincide
(e.g., van der Kruit 1988; K89), but the issue has yet to be
investigated conclusively.

The existence of a surface density threshold for star
formation is usually explained in terms of the Toomre criterion for
gravitational instability (Spitzer 1968;
Quirk 1972; Fall \& Efstathiou 1980; K89). 
Neither rotation nor pressure can stabilize a thin, gaseous,
differentially rotating disk if the Toomre Q parameter, 
\begin{equation}
Q(r) \equiv {c_s \kappa \over \pi G \Sigma_g},
\label{eq:Q}
\end{equation}
is less than unity (Safronov 1960; Toomre 1964; Goldreich \&
Lynden-Bell 1965; Binney \& Tremaine 1987); where the effective sound
speed $c_s$, the epicyclic frequency $\kappa$, and the gas surface
density $\Sigma_g$ all depend on radius $r$. K89 tested the
hypothesis that the Toomre criterion is responsible for the observed
cutoff in the SFR by measuring the ratio $\alpha
\equiv 1/Q$ of
the gas surface density $\Sigma_g$ to the critical surface density
\begin{equation}
\Sigma_c(r) \equiv {c_s \kappa \over \pi G} \equiv {\Sigma_g \over
\alpha} \equiv Q\Sigma_g
\end{equation}
for a sample of 15 spiral galaxies. Assuming a constant velocity
dispersion of $6~\kms$, he found that $\alpha \equiv \Sigma_g /
\Sigma_c \approx 0.5$\footnote{The Toomre criterion
for a gaseous disk depends on the sound speed. However, most
studies use the 1-D velocity dispersion $\sigma = c_s /
\sqrt{\gamma}$ instead, where $\gamma$ is the ratio of specific
heats. To correct for this, the values for
$\alpha$ quoted from these papers (K89, MK01, and HEB98) have
been multiplied by a factor $1/\sqrt{\gamma}$, assuming $\gamma =
5/3$ (which is incorrect if the gas is highly molecular or if the
velocity dispersion is dominated by turbulence). In addition,
the values quoted by K89 and HEB98 have been 
multiplied by a factor $\pi/3.36$ because they used the Toomre criterion
for a stellar disk instead of that for a gaseous disk.%
} at the truncation radius. This result was recently confirmed by MK01
who found from a sample 
of 32 well-studied nearby spiral galaxies that $\alpha = 0.53 \pm
0.2$. However, Hunter, Elmegreen, \& Baker (1998, hereafter HEB98)
measured $\alpha \approx 0.25$ for a sample of irregulars. 

In an important paper, Elmegreen \& Parravano (1994) 
proposed that star formation requires only the coexistence of two
thermal phases in pressure equilibrium and suggested that the fact
that Q becomes high beyond the optical edge is partly
coincidental. They argued that high values of the Q parameter prevent
star formation indirectly by removing the ability of the gas to form
the high pressure regions that are required for the existence of a
multiphase interstellar medium (ISM). The need for a cool phase was
also emphasized by 
Gerritsen \& Icke (1997), HEB98, Hunter, Elmegreen, \& van Woerden
(2001), Billett, Hunter, \& Elmegreen (2002), and Elmegreen (2002). 

There has been little discussion in the literature on the
appropriate value of the velocity dispersion $\sigma$ (but see
Ferguson et al.\ 1998), which is difficult to measure observationally
and is usually assumed to be independent of radius. K89 and MK01
assumed a constant value of $6~\kms$, while HEB98 used $9~\kms$ for
their sample of irregular galaxies. It is interesting to note that the
difference in the assumed velocity dispersion can account for the  
discrepancy in $\alpha$ between spirals and irregulars found by HEB98.
If HEB98 had assumed $\sigma = 6~\kms$, they would have
measured $\alpha \approx 0.37$ instead of 0.25, which would have
agreed with the result obtained by K89 and MK01 at the $0.8~\sigma$
level. 

Since the cold
phase has a temperature that is roughly two orders of magnitudes lower
than that of the warm phase ($T\sim 10^2~\K$ vs.\ $10^4~\K$), and thus
a thermal velocity dispersion that is smaller by a factor 10, it seems
reasonable to assume that the phase transition has important
consequences for the stability of the disk. Indeed, the importance of
cooling below $10^4~\K$ to the instability of disks 
has been recognized in other contexts, such as
the formation of the first objects from gas free of metals and dust (e.g.,
Corbelli \& Salpeter 1995; Corbelli, Galli, \& Palla 1997; Oh \&
Haiman 2002). However, since the presence of even modest amounts of dust
enhances the formation rate of molecular hydrogen by orders of
magnitude, the results of these studies are probably only valid for
the first generation of stars.

Here we investigate the importance of the transition from the warm
($T\sim 10^4~\K$) to the cold ($T\sim 10^2~\K$) interstellar phase to
the stability of disk galaxies using a model of a gaseous,
exponential disk embedded in a dark halo. The disk is
self-gravitating, contains metals and dust, and is illuminated by UV
radiation. The main conclusions are that 1) The transition to the cold
phase occurs at a surface density that agrees with the empirically derived
threshold surface density for star formation (e.g., Skillman 1987),
which supports the idea of Elmegreen \& Parravano (1994)
that the existence of a cold phase is critical for star formation;
2) The large decrease in the thermal velocity dispersion associated with the
transition to the cold phase triggers gravitational  
instability on a large range of scales.

Because the gas is thermally
unstable at intermediate temperatures (Field 1965) and because
self-shielding from the photodissociating UV-radiation provides a
positive feedback loop, the transition from the warm, atomic to the
cold, molecular phase is fairly sudden. Assuming hydrostatic
equilibrium (i.e., a self-gravitating disk), we find that the critical
surface density  is about $\Sigma_c \sim
3$--$10~\Msun\,\pc^{-2}$ ($N_{H,{\rm crit}} \sim 3$--$10\times
10^{20}~\cm^{-2}$), which
corresponds to a pressure $P/k \sim 10^2$--$10^3~\cm^{-3}\,\K$ and a
volume density $n_{H}\sim 10^{-2}$--$10^{-1}~\cm^{-3}$. The
threshold surface density is insensitive to the exact values of model
parameters such as the intensity of
the UV radiation, the metallicity, the turbulent pressure, and the
mass fraction in collisionless matter.

An important result obtained in this work
is that the phase transition 
triggers gravitational instability even in the presence of relatively strong
turbulence. The reason is that turbulent support increases the surface
density required for the phase transition by about as much as it
increases the velocity dispersion of the gas, thereby leaving the
$Q$ parameter \emph{at the radius of the phase transition} nearly
unchanged. Indeed, 
we find that the turbulent pressure needs to be more than 250 times
greater than 
the thermal pressure to prevent the phase transition from
triggering gravitational instability. For the atomic phase this would
correspond to a line of sight velocity dispersion greater than
$10(T/10^2~\K)^{0.5}~\kms$, 
which exceeds the values typical for the outer
parts of disks ($8~\kms$; e.g., Lo, Sargent, \& Young 1993;
Meurer et al.\ 1996). In
fact, there is little room for turbulence in 
the extended gas disks. If the gas is warm beyond the truncation
radius, as predicted by our model, the thermal velocity dispersion can
fully account for the observed \HI\ line widths. 

Although the hypothesis that the transition to the cold phase sets the
global star formation threshold in the outer parts of galaxies is
physically reasonable and, as we shall show, 
appears to be supported by observations, it is important to note that
there are other mechanisms that can trigger the formation of unstable,
cold clouds. For example, shocks from spiral density waves and
swing-amplifier instabilities can locally convert an otherwise
purely warm phase into a multiphase medium.\footnote{Note that
\emph{consequences} of star 
formation, such as expanding shells and turbulence generated by
supernovae, should not be taken into account when computing a global
star formation threshold even though they may be very important for
star formation in a multiphase ISM.} However, to prevent photoionization
from keeping the gas warm, these regions would also need to have surface
densities exceeding the threshold for the formation of a cold
phase and the phase transition would still trigger instability on
small scales.

This paper is organized as follows. Before discussing the model in detail in
\S\ref{sec:model}, we provide an intuitive derivation of the Toomre
criterion in \S\ref{sec:Toomre}. In this section we also derive an
equivalent instability criterion for the case when shear is more
important than the Coriolis effect in limiting the growth of
perturbations. Readers who want to get straight to  
the results can skip \S\ref{sec:Toomre} and read only the
first paragraph of \S\ref{sec:model}. In \S\ref{sec:cause} we
explore the physical cause of the existence of a critical radius for
gravitational instability. In
\S\ref{sec:recipes} scaling relations are presented for the
threshold surface density as a function of the metallicity, the
intensity of the UV radiation, the fraction of the pressure that is
non-thermal, and the fraction of the mass in gas. This section also
contains recipes for implementing star formation thresholds in
(semi-)analytic and numerical models of galaxy formation.
In \S\ref{sec:obs} we compare the 
model with observations and show that the predictions for the
threshold column densities and radii agree with the data. Furthermore,
it is shown that other predictions of the model, including a sharp
drop in the molecular fraction, are also supported by observations. It is also
demonstrated that the model correctly predicts the sizes of
stellar disks and their variation with disk scale length and
mass. Finally, \S\ref{sec:conclusions} contains a 
summary of the main conclusions.

\section{Gravitational instability in a rotating disk}
\label{sec:Toomre}
Although there are multiple derivations of the Toomre criterion (e.g.,
Binney \& Tremaine 1987) and the equivalent instability criterion for
shearing perturbations (e.g., Elmegreen 1993) in the literature, I
have been unable to find a previous derivation in terms of
timescales. The derivations are included here for pedagogical reasons.

Gravitational instability of a fluid requires that the
self-gravity of a perturbation exceeds its internal pressure. In terms
of timescales, a perturbation is gravitationally unstable if the local
dynamical timescale is smaller than the sound-crossing timescale
across the perturbation (or, equivalently, if the size of the
perturbation exceeds the local Jeans length).

The dynamical timescale for a perturbation of size $\lambda$ and
surface density $\Sigma$ within a thin disk is
\begin{equation}
t_{\rm dyn} \equiv \sqrt{\lambda \over G\Sigma} \sim \sqrt{{\lambda
\over a}}, 
\end{equation}
where $a$ is the gravitational acceleration. The sound crossing
timescale is 
\begin{equation}
t_{\rm sc} \equiv {\lambda \over c_s}.
\end{equation}
Hence, stability against gravitational collapse requires
\begin{equation}
\lambda < {c_s^2 \over G\Sigma}.
\label{eq:tdyntsc}
\end{equation}

Pressure is not the only restoring force in a rotating
disk. Unless the circular velocity relative to the center of the
galaxy scales as $v\propto r^{-1}$,
conservation of angular momentum 
forces a perturbation to rotate around its own center, thereby
providing centripetal support. The natural timescale for the internal
rotation of the perturbation is the epicyclic period $2\pi/\kappa$,
where $\kappa$ is the epicycle frequency,
\begin{equation}
\kappa^2 = 2\left ({v^2 \over r^2} + {v \over r}{dv \over dr}\right ).
\end{equation}
The equation of motion for a circular orbit can be obtained by
equating the dynamical timescale and the rotation period $t_{\rm
rot}$, and the criterion for instability is $t_{\rm dyn} < t_{\rm
rot}$. For the \emph{internal} rotation of a perturbation in a
rotating disk we have $t_{\rm rot}= 
2\pi/\kappa$, and stability against gravitational collapse thus
requires 
\begin{equation}
\lambda > {4\pi^2 G\Sigma \over \kappa^2} \equiv \lambda_{\rm crit}.
\label{eq:tdyntrot}
\end{equation}

Equations (\ref{eq:tdyntsc}) and (\ref{eq:tdyntrot}) cannot both be
satisfied if $c_s^2/G\Sigma < 4\pi^2 G\Sigma/\kappa^2$, i.e., if
$c_s\kappa/2\pi G\Sigma < 1$. As expected, this agrees with the Toomre
criterion, $Q<1$, to within a dimensionless factor of order unity (two). 
It is important to note that $Q<1$ does not imply instability to
perturbations of arbitrary wavelengths. Equations (\ref{eq:tdyntsc})
and (\ref{eq:tdyntrot}) indicate that perturbations with $\lambda <
c_s^2/G\Sigma$ are stabilized by pressure, while perturbations with
$\lambda > \lambda_{\rm crit}$ are stabilized by rotation. 

For a
perturbation of length $\lambda$, the exact dispersion relation (see,
e.g., equation [6-47] of Binney \& Tremaine 1987) can be
solved to give the critical value of $Q$, below which the perturbation is
unstable\footnote{A similar result can be derived from the relation
$1/t_{\rm dyn} = 1/t_{\rm sc} + 1/t_{\rm rot}$.},
\begin{equation}
Q_c(\lambda) = 2\sqrt{{\lambda \over \lambda_{\rm crit}} - \left
({\lambda \over \lambda_{\rm crit}}\right )^2}.
\label{eq:qcrit}
\end{equation}
As before, perturbations with $\lambda > \lambda_{\rm crit}$ are always
stable. For $\lambda \ll  
\lambda_{\rm crit}$ equation (\ref{eq:tdyntsc}) is recovered
exactly. Hence, for small wavelengths rotation is unimportant, and the
Toomre criterion asymptotes to the two-dimensional Jeans criterion. 

Besides forcing perturbations to rotate around their centers (the
Coriolis effect), galactic rotation can prevent the collapse of
perturbations through another mechanism: shear (e.g., Elmegreen 1993;
HEB98). If the disk is in 
differential rotation, then perturbations will be unable to collapse
unless the shearing timescale is longer than the dynamical 
timescale. The shear rate is given by Oort's constant $A$,
\begin{equation}
t_{\rm shr}^{-1} \sim \left | A\right |  
\equiv {1 \over 2} r \left |{d\Omega \over dr}\right |
= {1 \over 2}\left | {v \over r} - {dv \over dr}\right |.
\end{equation}
Stability against gravitational collapse requires $t_{\rm shr} <
t_{\rm dyn}$, i.e., 
\begin{equation}
\lambda > {G\Sigma \over  A^2}.
\label{eq:tdyntshear}
\end{equation}

Equations (\ref{eq:tdyntsc}) and (\ref{eq:tdyntshear}) cannot both be
satisfied if $Q_{\rm shr}$ is less than unity, where
\begin{equation}
Q_{\rm shr} \equiv {c_s A \over G\Sigma}.
\end{equation}
The ``$Q$ parameter'' for shearing perturbations is identical to the Toomre
parameter (equation \ref{eq:Q}), except that $\kappa$ is
replaced by $\pi A$. In particular, the 
critical surface density is again proportional to the velocity
dispersion. The epicycle frequency is compared with the factor $\pi A$
in Fig.~1 for two galaxy models (dot-dashed and dotted curves
respectively). 

The rotation curves of galaxies are observed to be flat in the outer
parts, i.e., $dv/dr \ll v/r$. In this limit $Q_{\rm
shear} \rightarrow Q \pi / 2\sqrt{2} \approx 1.11 Q$. If the rotation curve is 
rising, then $A$ can be much smaller than $\kappa$, indicating that
the Coriolis effect 
is more important than shear for limiting the growth of perturbations.
For all galaxy models tested, the
difference between the two $Q$ parameters at the transition to the
cold phase turned out to be small
compared to the drop in the velocity dispersion associated with this
phase transition (see e.g.\ Fig.~2). Thus, the presence of shear
does not change the conclusion that the phase transition causes
instability. In the following we will therefore generally only
consider the Toomre criterion,
noting that using the shear criterion instead would yield nearly identical
results.

\section{Model}
\label{sec:model}
To investigate the stability of disk galaxies, the model of Mo, Mao, \&
White (1998, hereafter MMW98) for an exponential disk embedded in a
dark halo, is combined with the photoionization
package CLOUDY (Ferland et al.\ 1998), which includes most of the
microphysics that is 
thought to be relevant for the ISM. For each radius CLOUDY is used to
compute the 
equilibrium temperature and ionization balance for a
range of densities, modeling the disk as a constant density
photo-dissociating region. From this grid of models, we 
then pick the density for which the disk
is in hydrostatic equilibrium. The methods used to compute the density
profile and the thermal and ionization balance are
discussed in detail in sections \ref{sec:densprof} and \ref{sec:thermal}
respectively. The results depend on the metallicity $Z$,
the UV radiation field $I$, and the dimensionless factor $f\equiv
f_g/f_{th}$, where $f_g$ is the fraction of
the mass in gas and $f_{th}$ is the ratio of the thermal pressure to
the total pressure. We choose $Z = 0.1 Z_\odot$, $I = 10^6$ hydrogen
ionizing photons $\cm^{-2}\,\s^{-1}$, and $f=1$ as our fiducial
parameters, a choice that is motivated in \S\ref{sec:pars}. In
\S\ref{sec:recipes} we show how the 
results scale with these parameters. In \S\ref{sec:cause} we
illustrate the results using two galaxy models: models HSB and LSB
have a mass $M_{200}$ and a dimensionless spin parameter $\lambda$ of 
$(10^{12}\,\Msun,0.05)$ and $(5\times 10^{10}\,\Msun,0.1)$
respectively (see \S\ref{sec:densprof} for further 
details). The rotation curves for these models are shown in Figure~1. 

The following cosmology will be adopted: total matter density 
$\Omega_m = 0.3$, vacuum energy density $\Omega_\Lambda = 0.7$, Hubble
constant $H_0 = 100~h~\kms\,\mpc^{-1}$, $h=0.65$. The helium mass
fraction is assumed to be $Y=0.24$.  

\subsection{Dynamics}
\label{sec:densprof}
MMW98 provide an analytic model for a self-gravitating disk embedded
in a dark halo based
on the following assumptions: (1) the disk contains a fraction $m_d$
of the total mass; (2) the disk contains a fraction $j_d$ of the total
angular momentum; (3) the disk is a thin, centrifugally supported
structure with an exponential density profile; (4) the disk is
embedded in a halo that initially had a NFW profile (Navarro, Frenk,
\& White 1997), but responded adiabatically to the assembly of the
disk (remaining spherical while it contracted).

NFW found that the following density profile provides a good fit to
the equilibrium density profiles of the cold dark matter (CDM) halos in
their $N$-body simulations,
\begin{equation}
{\rho(r) \over \rho_{\rm crit}} = {\delta_c \over (r/r_s)(1+r/r_s)^2},
\end{equation} 
where $\rho_{\rm crit} = 3H^2/8\pi G$ is the critical density,
$\delta_c$ is a characteristic density, and $r_s$ is a scale
radius. Defining the virial radius $r_{200}$ as the radius within
which the mean density is $200\rho_{\rm crit}$, it can be shown that
\begin{equation}
\delta_c = {200 \over 3}{c^3 \over \ln(1+c) - c/(1+c)},
\end{equation}
where $c$ the halo concentration factor, 
\begin{equation}
c \equiv r_{200}/r_s.
\end{equation}
The NFW profile for the density distribution of a dark halo is
specified by the redshift (which we will set to zero), the
cosmological parameters  
$(\Omega_m,\Omega_\Lambda,H_0)$, the concentration factor $c$, and
the mass $M_{200}$ (defined as the mass interior to $r_{200}$).

\begin{figure*}[t]
\begin{center} 
\epsscale{1.1}
\plotone{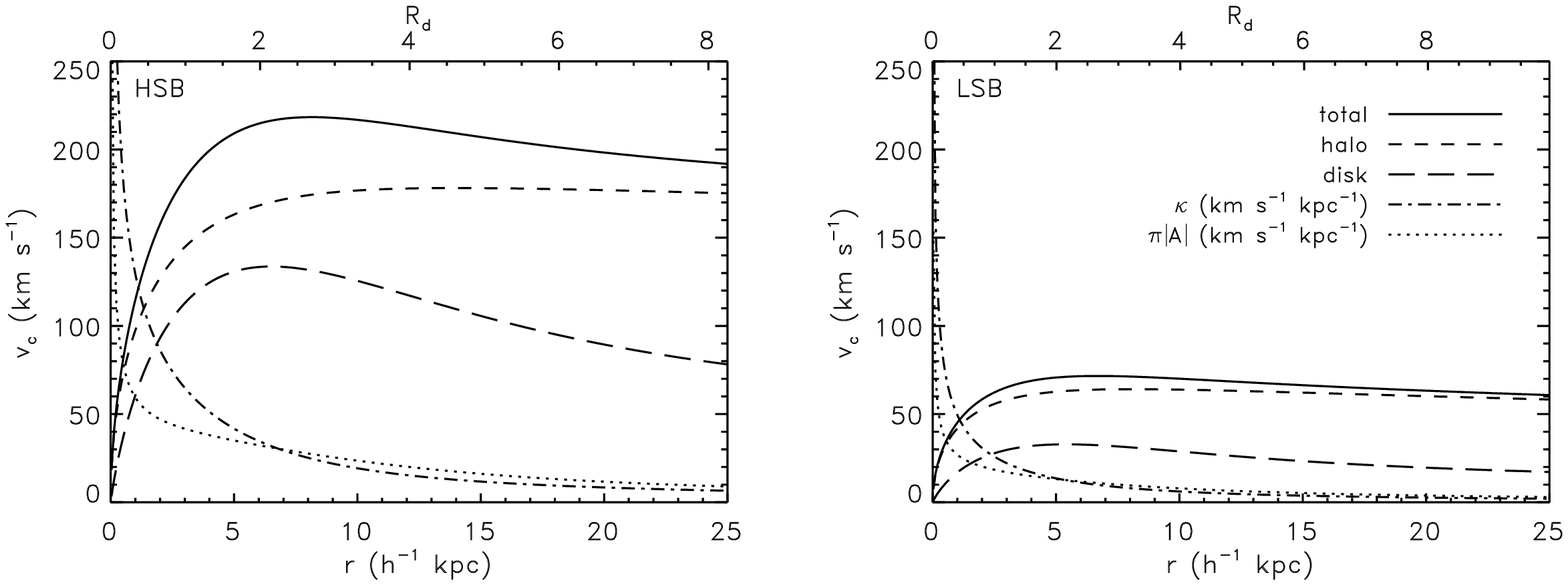}
\figcaption[f1.eps]{Circular velocity as a function of radius for
models HSB (\emph{left}) and LSB (\emph{right}), showing the
contributions of the halo (\emph{short-dashed curves}) and those of
the disk (\emph{long-dashed curves}). Also shown are the epicycle frequency
(\emph{dot-dashed curves}) and $\pi$ times Oort's constant $A$
(\emph{dotted curves}), both in $\kms\,\kpc^{-1}$.} 
\end{center}
\end{figure*}

The disk is assumed to have an exponential surface
density profile,
\begin{equation}
\Sigma = \Sigma_0 \exp(-r/R_d),
\label{eq:expdensprof}
\end{equation}
where $\Sigma_0 = m_d M_{200}/ 2\pi R_d^2$ and the disk scale length
is given by
\begin{equation}
R_d = {1 \over \sqrt{2}}\left ({j_d \over m_d}\right )\lambda r_{200}
f_c(c)^{-1/2} f_R(\lambda,c,m_d,j_d),
\end{equation}
where $f_c$ and $f_R$ are
dimensionless factors of order unity defined by equations (23) and
(32) of MMW98, and $\lambda$ is the dimensionless spin
parameter of the halo,
\begin{equation}
\lambda \equiv J\left |E\right |^{1/2}G^{-1}M_{200}^{-5/2},
\end{equation}
where $E$ is the total energy of the halo. The rotation
curve is computed assuming that the halo responds adiabatically to the slow
assembly of the disk as described in \S 2.3 of MMW98.  
For a given redshift and cosmology, the MMW98 disk model can thus be
fully specified by the parameters, 
$c$, $m_d$, $j_d$, $\lambda$, and $M_{200}$.

Most results will be presented for two model galaxies: a high and a low
surface brightness galaxy (models HSB and LSB respectively). We set
$z=0$, $c=10$, $m_d=0.05$, and $j_d = m_d$ for both models; MMW98
demonstrate that models with these parameter values (and distributions
of $\lambda$ and $M_{200}$ that fit CDM simulations) agree with the
observed Tully-Fisher relation (but see Borriello \& Salucci
2001). Models with $c\sim 10$ also provide 
good fits to the observed rotation curves of dwarf galaxies (van den
Bosch \& Swaters 2001). Model HSB has $M_{200} = 10^{12}~\Msun$
($v_{200} \approx 141~\kms$) and $\lambda = 0.05$, while model LSB has
$M_{200}=5\times 10^{10}~\Msun$ ($v_{200} \approx 52~\kms$) and
$\lambda = 0.1$. Hence, the disk scale length is 4.7 kpc for model HSB
and 3.9 kpc for model LSB.  The rotation curves for models HSB and LSB
are shown in Figure~1. The maximum circular velocity is $218~\kms$ for
model HSB and $72~\kms$ for model LSB. Figure~1 ({\emph{dot-dashed
curves}) shows the epicycle frequency. With this choice of parameters
models HSB and LSB are typical high and low surface brightness
galaxies in terms of their surface density profiles and rotation
curves. We have searched parameter space and checked that none of our
conclusions are specific to these particular models.

The disk is illuminated by an external radiation field. To compute the
thermal and ionization structure numerically, we need to specify the
full density profile $\rho(r,z)$. For an isothermal, exponential disk,
this distribution has been computed by Spitzer (1942). However, since
the radial variation in the temperature is one of the key ingredients
of the present work, we cannot assume isothermality. 

To simplify the numerics, the vertical density stratification is
neglected in the radiative transfer calculation, i.e., we assume
that the density varies smoothly with radius, but is a step function
in the $z$-direction. This assumption is reasonable because the column
density weighted density,
\begin{equation} 
\left <\rho(r) \right >_N \equiv {1 \over \Sigma(r)} \int \rho(r,z)^2 dz,
\end{equation}
which is the relevant density
for the radiative transfer calculation, is close to 
the midplane density for realistic profiles. In other words, in the
region that contributes significantly to the surface density, the
density is of order the midplane density. For example, for an
isothermal, exponential disk the midplane density is just 1.5 times
the column density weighted density.

The density and the column density can be related using the following
argument. In (local) hydrostatic equilibrium the
thickness of the disk, i.e., the size of the region over which the
density is of order the characteristic density, is of order the local
Jeans length $L_J$. Thus, the column density is of order the `Jeans
column density' (see Schaye 2001a for a derivation),
\begin{eqnarray}
N_{H,J} & \equiv & n_H L_J, \nonumber \\ 
 &=& \left ({\gamma k \over \mu m_H^2 G}\right )^{1/2} 
 (1-Y)^{1/2} f^{1/2} n_H^{1/2} T^{1/2}, \nonumber \\
&\approx & 3.06 \times 10^{21}~\cm^{-2}~ \mu^{-1/2} f^{1/2}
\left ({n_H \over 1~\cm^{-3}}\right )^{1/2} T_4^{1/2},
\label{eq:NJ} 
\end{eqnarray}
where $\gamma$ is the ratio of specific heats, $\mu$ is the mean
particle weight in units of the hydrogen mass 
$m_H$, $n_H$ is the hydrogen number density, $T \equiv T_4 \times
10^4~\K$, and $f \equiv f_g /f_{th}$ where $f_g$ is the fraction of
the mass in gas and $f_{th}$ is the ratio of the thermal pressure to
the total pressure. In the appendix it is 
demonstrated that for an isothermal, purely gaseous disk equation
(\ref{eq:NJ}) agrees with the exact solution to within 2 percent.

The hydrogen column density can be converted into a total gas surface
density using
\begin{equation}
\Sigma_g \approx 1.05 ~M_\odot\,\pc^2 ~\left ({N_H \over
10^{20}~\cm^{-2}}\right ) \left ({1-Y \over 0.76}\right )^{-1}.
\label{eq:sigma-NH}
\end{equation} 
For fixed $f_g$, there is a one-to-one relation between the
total pressure and the total hydrogen column density:
\begin{eqnarray}
P &=& {n_H kT \over \mu (1-Y) f_{th}}, \nonumber \\
&=& {m_H^2 G \over f_g\gamma (1-Y)^2} N_H^2, \nonumber \\
&\approx& {k \over f_g} \left ({N_H \over 2.67\times
10^{19}~\cm^{-2}}\right )^2 ~\cm^{-3}\,\K ,
\label{eq:pressure}
\end{eqnarray}
where we used equation (\ref{eq:NJ}) to go from the first to the second
equality. Hence, any critical column/surface density can be converted
into an equivalent critical pressure.

Note that if a cold phase is present, the situation is likely more
complicated than our model suggests. The presence 
of a cold phase leads to star formation and thus
turbulence, and the
amount of turbulent support may be different for the different
phases. More importantly, feedback from star formation produces a third,
hot phase and large variations in the local radiation field. The
model used here ignores these and other complications and can
therefore not reliably predict the structure of a multiphase ISM. What
it can do, however, is predict what is relevant for the present study:
the column density or, equivalently, the pressure at which the
transition to the cold phase occurs.

\subsection{Thermal and ionization balance}
\label{sec:thermal}

The thermal and ionization structure of the disk is computed
using the
publicly available photoionization package
CLOUDY (version 94; Ferland et al.\ 1998, Ferland 2000), modeling the
disk as a plane-parallel 
slab of constant density and assuming both thermal and ionization
equilibrium. Illumination from two sides is approximated by doubling
the column densities computed for a model illuminated from one
side, which is reasonable if the disk is strongly self-shielded. 
The heating and cooling rates are computed self-consistently from the
assumed incident continuum, which is described in \S\ref{sec:pars}.

For a given total hydrogen column density
$N_{H}$, density $n_H$, incident radiation field $I$,
metallicity $Z$, and dust-to-metals ratio, CLOUDY is used to
compute the column densities in neutral hydrogen $N_{HI}$ and molecular
hydrogen $N_{H_2}$, the temperature $T$, and the mean
particle mass $\mu m_H$ ($T$ and $\mu$ are first computed using a multi-zone
model of a photo-dissociation region, but are then averaged over disk height
$z$). CLOUDY contains nearly all microphysical processes 
that are thought to be relevant for the ISM and has been
tested extensively on a wide variety of problems. The reader is
referred to the
online\footnote{\texttt{http://www.nublado.org}.}
documentation for details.

The disk model of MMW98 provides us with the surface density as a
function of radius. Since we do not know a priori the density for which the
disk is in hydrostatic equilibrium (i.e., eq.\ [\ref{eq:NJ}] is
satisfied), we use the two-step procedure of Schaye (2001b)
to compute the structure of the disk. First, CLOUDY is used to compute
$T$ and $\mu$ for a grid of 
$(N_H,n_H)$ models. Second, for each radius (i.e., for each value of
$N_H$) those solutions
$(N_H,n_H,T(N_H,n_H),\mu(N_H,n_H))$ are selected for which equation
(\ref{eq:NJ}) is satisfied and which are stable ($dP/dn_H > 0$). If
there are two stable solutions for a fixed $N_H$, then the low
temperature solution is selected because we are interested in the smallest
column density for which the cold phase exists. Since the range of $N_H$
for which two solutions are possible is small for all the models that
were tested, picking the high 
temperature solution instead would not change the results
significantly. Note
that if $N_H$ is large enough that only the low temperature solution is
possible, this does not mean that there is no warm
phase, but merely that the fraction of gas in the warm phase, which
always exists at sufficiently large scale heights, is small compared
to that in the cold phase.

In summary, for each radius (i.e., surface density), we first compute
the equilibrium temperature and ionization balance for a range of
densities and then we pick the lowest temperature solution for which
the disk is thermally stable and in hydrostatic equilibrium.

\subsection{Fiducial parameter values}
\label{sec:pars}
The parameters for the MMW98 disk model [$c=10$, $m_d=0.05$,
$j_d=m_d$, $(M_{200},\lambda)_{\rm HSB} = (10^{12}\,\Msun,0.05)$, 
$(M_{200},\lambda)_{\rm LSB} = (5\times 10^{10}\,\Msun,0.1)$] have  
already been discussed in \S\ref{sec:densprof}. In this section we
will motivate our fiducial values for the parameter $f$ (recall that
$f\equiv f_g/f_{th}$ where $f_g$ is the fraction of the mass in gas and
$f_{th}$ is the ratio of the thermal pressure to the total pressure),
the metallicity $Z$, the intensity of the UV radiation $I$, and the 
dust-to-metals ratio. In \S\ref{sec:recipes} we will see how
the results change if different parameter values are
used. Fortunately, the star formation threshold turns out to be
insensitive to small (an order of magnitude or less) variations in these
parameter values.

It is unclear how the factor $f$ can be measured, but it seems
reasonable to assume that $f\sim 1$ beyond the critical radius for
star formation, since both the mass fraction in gas and the ratio of
the thermal to total pressure are likely to be close to unity in the absence
of star formation. For this reason we choose $f=1$ as our fiducial
value. 

The metallicity is measurable, although measurements based on emission
lines are generally only possible in regions of ongoing star
formation, while absorption studies require the good fortune of a
bright background source. Emission line studies of \HII\ regions
indicate that the metallicity is generally 
of order 10 percent solar near the optical edge (e.g., Ferguson et
al.\ 1998a; Henry \& Worthey 1999). Studies of damped 
Ly$\alpha$ absorption lines give similar values, although the scatter
is large (e.g., Pettini et al.\ 1999). We will use $Z=0.1~Z_\odot$ as
our fiducial value.

Comparisons of the relative abundance of refractory and non-refractory
elements in damped Ly$\alpha$ systems indicate that the
dust-to-metals ratio is about half that of the Galactic value (e.g.,
Pettini et al.\ 1997; Vladilo 1998), and we will therefore use this as
our fiducial value. The default dust composition of CLOUDY is used,
i.e., a mixture of silicates and graphites with ISM
properties. Depletion of metals onto dust grains is taken into
account, using CLOUDY's default depletion factors.

\begin{figure*}[t]
\begin{center} 
\epsscale{1.1}
\plotone{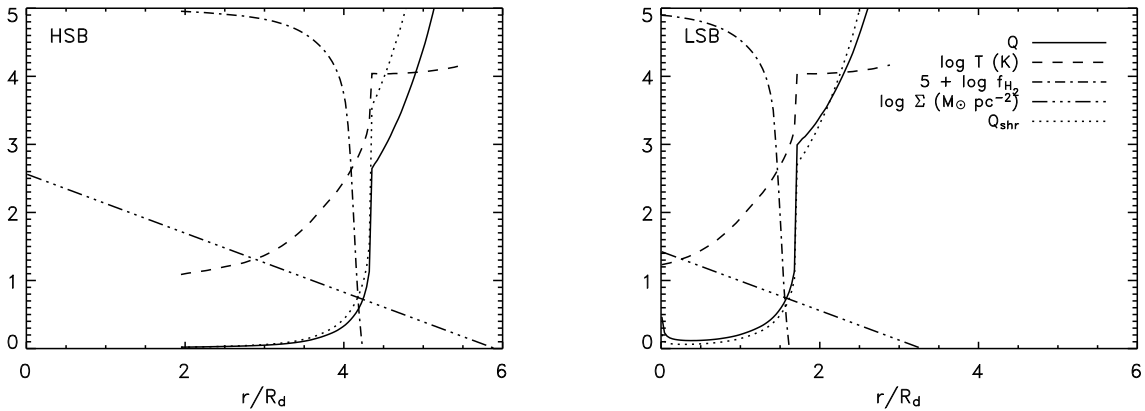}
\figcaption[f2.eps]{The Toomre $Q$ parameter (\emph{solid curves}), the
temperature $\log T$ (\emph{dashed curves}), the molecular fraction
$5+\log f_{H_2}$ 
(\emph{dot-dashed} curves), the surface density $\log\Sigma (\Msun\,\pc^{-2})$
(\emph{triple-dot--dashed curves}), and the $Q$ parameter for shearing
pertubations 
(\emph{dotted curve}) are all plotted as a function of radius 
for models HSB (\emph{left}) and LSB (\emph{right}). The sudden drop in
the Q values coincides with (and is caused by) a similar drop in the
temperature, and a sharp increase in the molecular fraction. The
transition to the cold phase, which coincides with 
the onset of gravitational instability ($Q<1$), occurs at a fixed
surface density. Note that the models become unrealistic shortwards of
the critical radius, where feedback from star formation will increase
the UV field, the metallicity, and the turbulent pressure.} 
\end{center}
\end{figure*}

\begin{figure*}[t]
\begin{center} 
\epsscale{1.1}
\plotone{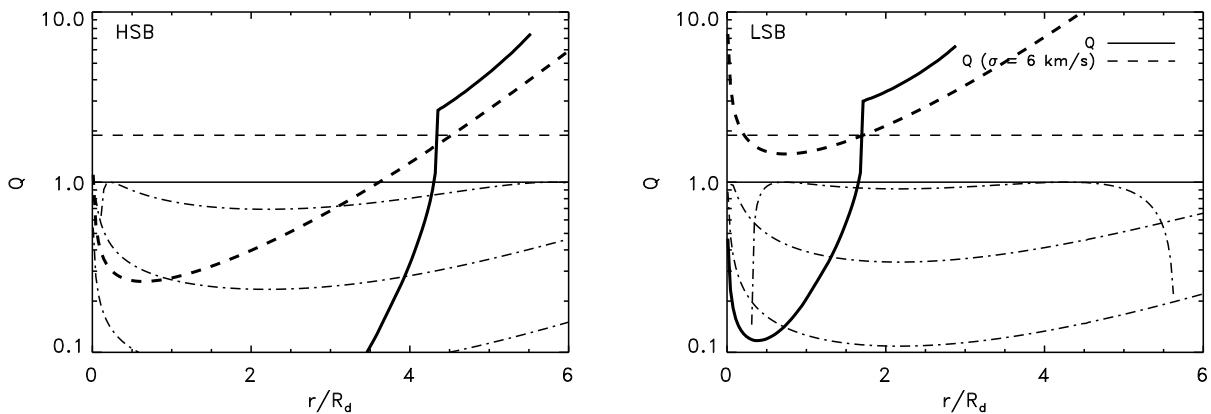}
\figcaption[f3.eps]{The Toomre $Q$ parameter as a function of radius
(\emph{thick solid curves}) for models HSB (\emph{left}) and LSB
  (\emph{right}). The 
solid horizontal line indicates $Q=1$. The sudden drop in the Q-value
is caused by the transition to the cold phase. The dotted curves
indicate the $Q$ values below which the disk in unstable to
perturbations with wavelengths of (\emph{top}) $10^3$, (\emph{middle})
$10^2$, and (\emph{bottom}) 10~pc. Assuming a constant velocity
dispersion of $6~\kms$ leads to large errors in the
Q-parameter (\emph{thick dashed curves}). The transition to the cold
phase, which causes 
gravitational instability ($Q<1$), occurs at about the radius where 
$Q(\sigma = 6~\kms) = 1/0.53$ (\emph{intersection of thick dashed curves
with horizontal dashed lines}), the empirical star formation threshold
of MK01. This 
implies that the hypothesis that the star formation threshold is
associated with the transition to the cold phase, is consistent with the
observations. Note that the
models become unrealistic shortwards of the critical radius, where
feedback from star formation will increase the UV field, the
metallicity, and the turbulent pressure.} 
\end{center}
\end{figure*}

The most uncertain parameter is undoubtedly the intensity of the
UV radiation. Measurements of the present-day intensity of the
extragalactic UV background provide a
strong lower limit of $I \sim 10^{4.5}$ hydrogen ionizing photos
$\cm^{-2}\,\s^{-1}$ 
(ionization rate $\Gamma \sim 10^{-14}~\s^{-1}$) with an uncertainty
of a factor of a few (Scott et al.\ 2002 and references
therein). 
However, the UV radiation illuminating the thick gas disk
just beyond the critical radius is likely to be significantly more
intense than this, and could well show strong spatial fluctuations. In
addition to the extragalactic UV background (for which we assume the
spectral shape of the model of Haardt \& Madau 2001, which includes
the extragalactic X-ray background, and the normalization of Scott et
al.\ 2002), we use a 
component with the spectral shape of the unextinguished local
interstellar radiation field of Black (1987) (using the CLOUDY command
``table ISM'') so that the total, i.e., extragalactic plus local,
intensity of the UV radiation above 1 Rydberg is $I~{\rm
photons}~\cm^{-2}\,\s^{-1}$. Finally, we add two more
components: the cosmic microwave background and a cosmic ray density
of $2\times 10^{-9}~\cm^{-3}$. However, the addition of these last two
components does not have a significant effect on the star formation
threshold. 

The model of 
Bland-Hawthorn \& Maloney (1999; 2002, see their Fig.~3)
for the UV radiation field of the 
Galaxy, which includes contributions from the bulge, disk, halo, and
the cosmic background, predicts an intensity of order $10^6~{\rm
ph}~\cm^{-2}\,\s^{-1}$ at the truncation radius of the Galaxy
($\approx 12~\kpc$, Freudenreich 1998), and we take this as our
fiducial value. For LSB galaxies and \HII\ regions beyond the critical
radius a lower value may be more appropriate. In
\S\ref{sec:recipes} it is shown that using just the extragalactic
radiation field, reduces the critical surface density only by about a
factor two.

Apart from the normalization, the shape of the spectrum also affects
the surface density at which the phase transition takes place. A
harder spectrum increases the heating rates, which increases the
critical surface density. To test the sensitivity to the assumed
spectral hardness, we computed a model using the spectral shape of the
Haardt \& Madau (2001) model for the extragalactic background (which
has a softness parameter $S\equiv \Gamma_{\rm HI}/\Gamma_{\rm HeII}
\approx 5.6\times 10^2$, whereas the spectrum of Black 1987 has $S
\approx 2.7\times 10^3$) for the local UV radiation, leaving the
normalization unchanged. Again, the difference in the critical surface
density is only a factor of two.

\section{The physical cause of star formation thresholds}
\label{sec:cause}

The solid curves in figure~2 show the Toomre $Q$ parameter as a
function of radius for models HSB (\emph{left}) and LSB
(\emph{right}). At the critical radius $r_c$ the $Q$ parameters drop 
sharply from $Q>2$ at $r>r_c$ to values smaller than unity at $r<
r_c$ ($r_c\approx 4.3 R_d$ for model HSB and $\approx 1.7 R_d$ for model
LSB). This sudden decrease is associated with a similar drop in the
temperature (\emph{dashed curves}) from $T\approx 10^4~\K$ to $< 10^3~\K$
and with a sharp increase in the molecular hydrogen fraction (the
dot-dashed curves in figure~2 show $5 + \log f_{H_2}$) from $f_{H_2}
\ll 10^{-3}$ to more than $10^{-3}$. 

As discussed in \S\ref{sec:Toomre}, besides the Coriolis effect,
which the Toomre criterion takes into account, shear is another
consequence of galactic rotation which can prevent the growth of
perturbations. The dotted curves in figure~2 show that, as expected
(see \S\ref{sec:Toomre}), the instability parameter for shearing
perturbations, $Q_{\rm shr}$, shows a similar behavior as the Toomre
$Q$ parameter, indicating that shear cannot stabilize the cold phase.

The fact that $Q$ is significantly greater than unity for $r>r_c$ and
then drops sharply to $Q<1$ at $r\approx r_c$ implies that it is the
transition to the cold phase that causes gravitational instability and
not vice versa. Note that since both the surface density
(Fig.~2, \emph{triple-dot--dashed curves}) and the epicycle frequency
(Fig.~1, \emph{dot-dashed curves}) vary smoothly across $r_c$, it must be
the decrease in the sound speed associated with the phase transition
that causes the drop in $Q$. 

As discussed in \S\ref{sec:Toomre}, $Q<1$ only implies
instability to perturbations of size $\lambda_{\rm crit}/2$, which is
typically $\ga \kpc$ in the outer disk. Instability on smaller scales
requires smaller values of $Q$ (see eq.~[\ref{eq:qcrit}]). Figure~3
demonstrates that 
perturbations with $\lambda \ll \lambda_{\rm crit}$ 
do become unstable in the cold phase. The dash-dotted curves indicate the
$Q$ thresholds required for the instability of perturbations with
wavelengths of (\emph{top to bottom}) $10^3$, $10^2$, and 10~pc
respectively. The sharpness of the drop in the $Q$ value associated
with the phase transition causes fluctuations of 1~kpc and $10^2$~pc to
become unstable at nearly the same surface density.

The transition to the cold phase coincides with a large
increase in the molecular fraction. This is because the density, and
thus the \Htwo\ formation rate, must increase as the temperature
decreases to prevent a drop in the internal pressure.
Furthermore, the higher molecular fraction increases the cooling rate,
providing a positive feedback loop. Other reasons why the transition
to the cold phase 
is sudden are that the gas is thermally unstable at intermediate
temperatures ($T\sim 10^3 - 10^4~\K$) and that the increase in the
molecular fraction enhances the self-shielding from photodissociating
radiation, thereby providing another positive feedback loop.

After the phase transition the gas is gravitationally unstable and
will fragment into clouds with a higher column density 
than the azimuthally averaged column density used in the models. This
will make self-shielding more effective and will thus lead to a further
decrease in the 
temperature and a further increase in the molecular fraction. This
self-reinforcing process of cooling and collapse will not continue
indefinitely, as the clouds will eventually become opaque to their own
cooling radiation. Observations indicate that molecular clouds
do form stars and it therefore seems plausible to associate
the transition to the cold phase, which we have shown to trigger
gravitational instability on a large range of scales, with the critical
radius for star formation. In \S\ref{sec:obs} we will show that the
observations support this hypothesis.

Equation~(\ref{eq:NJ}), which assumed local hydrostatic equilibrium,
shows that if the temperature $T$, the mean 
molecular weight, and the factor $f$ ($\equiv f_g/f_{th}$, where $f_g$
is the fraction of the mass in gas and $f_{th}$ is the ratio of the
thermal to the total pressure) are constant, as is roughly the case
beyond the critical radius where only the warm phase is present, then
there is a one to one relation between density and column density or,
equivalently, pressure. For fixed $T$, $f$, metallicity $Z$, and UV
radiation field $I$, the existence of a cold phase depends on the
density and the column density. Because the latter two are related
one-to-one, the phase transition occurs at a fixed surface density 
($\log N_{H,{\rm crit}} \approx 20.75$ for our fiducial
parameters). Hence, there exists a threshold surface density for star
formation, which does not depend on the rotation curve of the galaxy.
The difference in the critical radii between the two models can
therefore be explained in terms of their surface density profiles
(Fig.~2, \emph{triple-dot--dashed curves}). Low surface brightness
galaxies have lower 
surface densities, and therefore smaller optical radii (relative to
their disk scale lengths). Note that in reality the situation could be
more complicated if the values of the model parameters ($f$, $Z$, $I$)
differ systematically between the outer disks of HSB and LSB galaxies.

Although the calculation of the radius within which the disk is 
unstable to star formation is robust, the detailed
predictions of the models may be unrealistic for $r<r_c$. The model
effectively predicts its own demise: within the critical radius
feedback from star formation will modify 
the thermal and ionization structure of the disk and will generate  
turbulence, possibly
leading to self-regulation of the SFR such that $Q\approx
1$. The energy injected by young stars and supernovae 
will also convert gas from the cold phase into the warm and/or
a third (hot) phase, and may introduce large fluctuations
in the UV radiation. The models presented here may therefore
be inadequate for studying star formation in the multiphase ISM.
However, we emphasize that the prediction of widespread star formation
at $r<r_c$ is robust, since the conditions that invalidate the model
are \emph{consequences} of this prediction. 

It should, however, be noted that there are physical processes that
have not been included in the models, which could result in the
formation of high column density clouds in an otherwise warm
phase. Examples are swing-amplifier instabilities, gravity-driven 
turbulence, infalling gas clouds, and spiral density waves.
However, to form stars a cold phase is 
still required and consequently regions unstable to star formation
should still have surface densities that exceed the threshold value.
Hence, the local surface density threshold for star formation would remain
unchanged, but the conclusion that the phase transition triggers
gravitational instability may no longer be correct, except
for small-scale perturbations. Regardless of what
comes first, such processes break the axisymmetry
that was assumed in the models, and will thus break the 
one-to-one relationship between surface density and radius. This
would complicate the interpretation of azimuthally smoothed
observations, but it would not be in conflict with our criterion for star
formation, which is a local threshold.

\subsection{The effects of turbulence}
\label{sec:turbulence}

Turbulence both prevents and promotes gravitational instability. It
inhibits global collapse because it provides an additional source of
pressure, but it can trigger local instability by creating density
fluctuations. We will discuss each of these effects in turn.

Since the warm phase is gravitationally stable even in the absence of
turbulence, increasing the amount of turbulent support will not change
the conclusion that the presence of a cold phase is necessary for
small-scale gravitational instability. However, if the turbulence is
sufficiently strong, the presence of a cold phase may no longer be
sufficient for 
instability. Turbulent support affects the $Q$ parameter at the phase
transition in two opposing ways. First, turbulent pressure increases
the stability of the disk, approximately as\footnote{This scaling is
only correct if magnetic pressure is negligible.} $Q \propto
f_{th}^{-0.5}$ (recall that $f_{th} \equiv P_{th} / P \propto
\sigma_{th}^2 / \sigma^2$). Second, it increases the critical surface
density for the transition to the cold phase (because it decreases the
equilibrium density corresponding to a fixed surface density; see
eq.~[\ref{eq:NJ}]), which leads to a decrease in the
$Q$ parameter at the phase 
transition. In the next section we will show that this critical surface
density
scales approximately as $\Sigma_{\rm crit} \propto f_{th}^{-0.3}$. Thus,
increasing the turbulence shifts the phase transition to higher
surface densities (i.e., smaller radii) and increases the Q-value at
this radius, approximately as $Q\propto f_{th}^{-0.5}f_{th}^{0.3} =
f_{th}^{-0.2}$. Given that $Q$ is very small within the critical
radius for $f_{th}=1$ (see, e.g.,  Fig.~2), it is 
clear that extremely large turbulent pressures are required to shift
the instability to radii smaller than the radius of the phase
transition.

Indeed, in our models the phase transition triggers instability for
values of $f_{th}$ as low as $1/250$, which corresponds to a 
line of sight velocity dispersion of $10(T/10^2~\K)^{0.5}~\kms$ (for
model LSB the derived critical surface densities 
exceed the central surface density if $f_{th} \ll 10^{-1}$). Higher
turbulent pressures are ruled out as they would result in line of
sight velocity dispersions that are significantly greater than the
$8~\kms$ that is 
typical for the extended \HI\ disks of galaxies (e.g., Lo et
al.\ 1993; Meurer et al.\ 1996). 

Since turbulence locally compresses gas it can
also induce instability. If the compression is sufficiently strong and
happens on a timescale short compared with the cooling time, then
the compressed gas could become gravitationally unstable while it is
still in the warm phase. This situation is common in the
multiphase ISM, where feedback from star formation is known to generate
supersonic turbulence. As long as the turbulence is generated by feedback
from star formation, we can ignore it for our purposes because we are only
concerned with the validity of the models in the absence of star
formation. There are, however, other processes capable of generating
turbulence, such as swing-amplifier instabilities, shocks from
spiral density waves, and infalling gas clouds. 

Although turbulence likely plays a central role in the multiphase ISM,
there is in fact very little room for turbulence in the outer disk,
because thermal motions in the gas heated by (extragalactic) UV
radiation can by itself account for the observed widths of the \HI\
lines. Indeed, proposed sources of turbulence in the extended gas disk
appear unable to account for the observed velocity dispersion.
 Sellwood \& Balbus (1999) estimate that MHD turbulence
may result in a turbulent velocity dispersion of $6~\kms$ (depending
on the unknown strength of the magnetic field) and simulations
by Wada, Meurer, \& Norman (2002) show that gravity driven
turbulence yields a velocity dispersion of only $2-3~\kms$ in the
outer disk. Thus, it appears likely that the turbulence in the
extended gas disk, where there is no feedback from star formation, is
mostly subsonic.

\subsection{On the applicability of the Toomre criterion}
\label{sec:validity}
In this section we further diminish
the role of the Toomre criterion for star formation in the outer parts
of galaxies by showing that it is not
applicable to the warm extended gas disk and that it reduces to the Jeans
criterion for the cold phase. 

From equation (\ref{eq:NJ}) it can be seen that the disk
thickness is about,
\begin{equation}
L_J \sim 4.4~\kpc~\left ({N_H \over 10^{20.75}}\right )^{-1} f T_4.
\label{eq:LJ}
\end{equation}
Thus, the transition from the cold to the warm phase results in a
large increase in the disk scale height, and the disk is predicted to
flare beyond the critical radius. However, in reality the flaring may
be much less pronounced than predicted by equation (\ref{eq:LJ})
because the increase in
$T$ will be partially offset by a decrease in $f_g$ (and
thus $f$), because the effect of the dark halo may not be
negligible for the warm, outer disk (e.g., Maloney 1993; Olling
1995). Moreover, if star formation is indeed ubiquitous when a cold
phase is present, then the resulting turbulent pressure would decrease
$f_{th}$ (and thus increase $f$) within the critical radius. 

\begin{figure}
\begin{center}
\epsscale{1.1}
\plotone{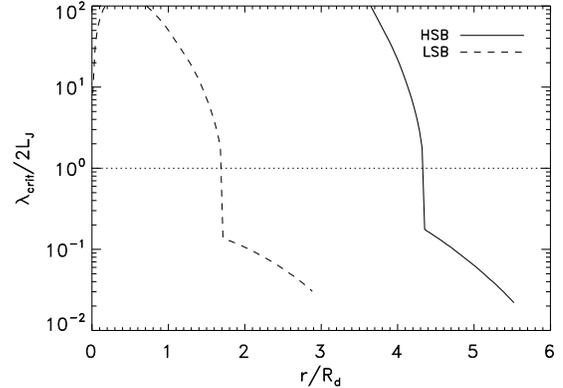}
\figcaption[f4.eps]{The ratio of the most unstable wavelength
$\lambda_{\rm crit}/2$ (unstable for $Q<1$) relative to the disk
thickness ($\sim L_J$) as a 
function of radius. Longward of the critical radius for star
formation, $\lambda_{\rm crit}$ is much smaller than the disk
thickness, thereby invalidating the assumption that the disk is thin.} 
\end{center}
\end{figure}

Figure~4 demonstrates that the Toomre analysis is invalid for the
warm, outer disk because the assumption that the disk is thin breaks down.  The
figure shows the ratio of the most unstable wavelength $\lambda_{\rm
crit}/2$ (i.e., the wavelength of perturbations for which $Q_c = 1$;
see eq.~[\ref{eq:qcrit}]) to the disk thickness $L_J$ (assuming
$f=1$), as a function of radius.  Outside the critical radius the
ratio $\lambda_{\rm crit}/2L_J$ is much smaller than unity, implying
that the disk cannot be considered to be thin when investigating the
stability of perturbations of size $\lambda\sim \lambda_{\rm
crit}/2$. Since a thick
disk is more stable than a thin disk (Romeo 1992), this reinforces the
conclusion that the warm phase is gravitationally stable. The
transition to the cold phase results in a large decrease in the disk
scale height, and the assumption that the disk is thin does hold
shortwards of the critical radius ($\lambda_{\rm crit}$ depends only
on $\Sigma$ and $\kappa$ [see eq.~[\ref{eq:tdyntrot})] and therefore
varies smoothly across the critical radius).

The assumption that the disk is thin always becomes invalid for
sufficiently small wavelengths. Figure~3 shows that the phase
transition triggers instability on a large range of length scales.
For
perturbations with $\lambda \ll \lambda_{\rm crit}$, rotation 
is unimportant, and the Toomre criterion is equivalent to the Jeans
criterion (see \S\ref{sec:Toomre}). 
Note, however, that since $L_J
\sim 10^2~\pc$ in the cold phase (eq.~[\ref{eq:LJ}]), the assumption
that the disk is thin
is only marginally valid for the smallest wavelengths that are
predicted to be unstable. 

Thus, for models HSB and LSB rotation has no effect on the star
formation threshold in the outer disk. However, since the epicycle
frequency diverges at $r=0$, rotation will formally always stabilize
the innermost parts of the disk. For example, rotation is responsible
for the upturn in the $Q$ parameter at $r<0.3 R_d$ that can be seen in
figures 2 and 3 for model LSB. This effect is a direct consequence of
the assumption that the galaxy can be described by a thin exponential
disk all the way 
to $r=0$. In reality this assumption breaks down and the epicycle
frequency does not diverge. We have checked a wide variety of galaxy
models and find that rotation can only stabilize the cold phase in the
innermost part of the disk, $r\ll R_d$, where the epicycle frequency
starts to diverge. The model predictions for the critical radius are
therefore robust as long as $r_c \ga R_d$, as is the case for
observed disk galaxies. Note, however, that if one is interested in
galaxies that are close to being completely dark [$\Sigma(r=0) \approx
\Sigma_{\rm crit}$], then any model that is based on a thin,
exponential disk will give misleading results.

\section{Star formation recipes}
\label{sec:recipes}
The critical surface density or, equivalently, the critical pressure
above which a cold phase exists, depends on the 
factor $f$ ($f \equiv f_g/f_{th}$, where $f_g$ is the mass fraction in
gas and $f_{th}$ is the ratio of the thermal to total pressure), the
metallicity $Z$, the intensity of the UV radiation $I$, and the
dust-to-metals ratio. Because the critical radius for star
formation in the outer disk is set by the transition to the cold phase,
the star formation threshold is insensitive to the rotation
curve of the galaxy. In this section we will
present scaling relations for the critical surface density for star
formation, which can be used in (semi-)analytic models of galaxy
formation and evolution. Finally, we will present critical
\emph{volume} densities, which can be used to implement star formation
thresholds in hydrodynamical simulations.

The sign of the dependence of the critical surface density on
the various parameters is easy
to predict: the phase transition occurs
at a higher column density if the fraction of the mass in a
collisionless component is lower ($f_g$ higher), if the non-thermal
pressure is higher ($f_{th}$ lower), if the metallicity and dust
content are lower ($Z$ lower), and if the UV radiation is more intense
($I$ higher). The exact scaling relations are, however, difficult to compute
analytically. We will therefore simply provide fits to the
results of our numerical calculations.

The following empirical formula provides a satisfactory fit to the
column density at which the temperature is 500~K,
\begin{eqnarray}
\lefteqn{\log N_H(T=500\,\K) \approx 20.75} \\
&& + 0.29\log(f) + 0.0052\log^2(f) 
\nonumber \\
&& - 0.32 \log(Z/0.1 Z_\odot) - 0.047\log^2(Z/0.1 Z_\odot)
\nonumber \\
&& + 0.23 \log(I/10^6~\cm^{-2}\,\s^{-1}) + 
0.027\log^2(I/10^6~\cm^{-2}\,\s^{-1}) \nonumber .
\label{eq:t500}
\end{eqnarray}
The column density for which the molecular fraction reaches one part
in a thousand is fit well by the following formula,
\begin{eqnarray}
\lefteqn{\log N_H(f_{H_2}=10^{-3}) \approx 20.75} \\
&& + 0.31\log(f) + 0.0025\log^2(f) 
\nonumber \\
&& - 0.32 \log(Z/0.1 Z_\odot) - 0.051\log^2(Z/0.1 Z_\odot)
\nonumber \\
&& + 0.26 \log(I/10^6~\cm^{-2}\,\s^{-1}) + 
0.025\log^2(I/10^6~\cm^{-2}\,\s^{-1}) \nonumber .
\label{eq:fh2scaling}
\end{eqnarray}
Finally, the following formula provides a good fit to the
minimum column density for which the Toomre parameter equals
unity
\begin{eqnarray}
\label{eq:sigmacq1}
\lefteqn{\log N_H(Q=1)\approx 20.68} \label{eq:qscaling} 
\\
&& + 0.28\log(f_g) + 0.020\log^2(f_g) 
\nonumber \\
&& - 0.35\log(f_{th}) + 0.030\log^2(f_{th}) 
\nonumber \\
&& - 0.30 \log(Z/0.1 Z_\odot) - 0.047\log^2(Z/0.1 Z_\odot)
\nonumber \\
&& + 0.22 \log(I/10^6~\cm^{-2}\,\s^{-1}) + 
0.041\log^2(I/10^6~\cm^{-2}\,\s^{-1}) \nonumber .
\end{eqnarray}

The formulas given above can be used to implement star formation
thresholds in (semi-)analytic models of galaxy formation and
evolution. The fits were determined by running the model described in
\S\ref{sec:model} for varying values of the parameters $f_g$, $f_{th}$,
$Z$, and $I$. Each parameter was first varied separately,
then the fits were optimized by varying multiple parameters
simultaneously. Tests show that the errors in the fits to the predicted
critical column 
densities are smaller than 0.25~dex over (at least) the range $f =
10^{-3} - 10^2$, $Z = 10^{-4}- 10~Z_\odot$, and $I = 10^4 - 10^8~{\rm
photons}~\cm^{-2}\,\s^{-1}$, with the exception of
$N_H(T=500~\K)$, which can be significantly in error for $Z \ga 3
Z_\odot$ because the large metal abundance causes the gas to cool
below 500~\K\ before the molecular fraction reaches $10^{-3}$ and
the disk becomes unstable. 

Although the above formulas fit the results of the models to within
0.25~dex, the systematic errors caused by the limitations of the model
may be somewhat larger than 0.25~dex. In particular, our simplified
prescription for hydrostatic equilibrium (see \S\ref{sec:densprof}), 
cannot be expected to predict the critical surface densities with an
accuracy better than a factor of a few. For example, the Jeans length
is often defined with a factor of $\sqrt{\pi}$ that was not included in
equation (\ref{eq:NJ}). The effect of including this factor is
equivalent to setting $f_g = \pi$; i.e., the normalization of the
three critical surface densities would be increased by about 0.15~dex.

Note that $N_H(Q=1)$  depends explicitly on $f_{th}$ because the
$Q$ parameter depends directly on the turbulent velocity dispersion. 
However, $N_H(Q=1)$ scales with $f_{th}$ in roughly the same way as
$N_H(T=500~\K)$ and $N_H(f_{H_2}=10^{-3})$.
Because the drop in the thermal velocity dispersion is large compared
with the variation in the epicycle frequency at the phase transition,
$N_H(Q=1)$ is very insensitive to $\kappa$. We have tested a 
wide variety of galaxy models\footnote{The following parameter ranges were
tested: $\lambda j_d/m_d = 0.01 - 0.3$, $c =
1 - 100$, $m_d = 0.005 - 1.0$, $M_{200} = 10^9$ to $5\times
10^{13}~\Msun$.} and find that equation 
(\ref{eq:sigmacq1}) works as long as $r_c \ga R_d$. As
discussed in \S\ref{sec:validity}, for $r \ll R_d$ the epicycle
frequency diverges and the innermost parts of disks are thus formally
stabilized by 
rotation. However, for $r\ll R_d$ the scale height is no longer 
small compared to the radius and the Toomre criterion is not
applicable. Furthermore, the centers of disk galaxies are generally
dominated by a bulge component. 

All three of the above formulas yield critical 
surface densities that depend only weakly on the values of the
parameters $f$, $Z$, and $I$. Typically, a parameter has to change by
a factor $\sim 10^3$ for the threshold surface density to change by a
factor 10. The reason for this insensitivity to the values of the
model parameters is hydrostatic equilibrium. For example, naively one
might think 
that an increase of the UV radiation by a factor of 10 would result in
a similar increase in the critical surface density. However, in
hydrostatic equilibrium the volume density scales as the surface
density squared (see eq.~[\ref{eq:NJ}]) and because the cooling
rate and the formation rate of molecular hydrogen increase with
increasing density, the increase in the critical surface density is
only about a factor of 2. It is important to note that if we had not
assumed the disk to be self-gravitating (i.e., in approximate
hydrostatic equilibrium), then agreement between the critical surface
density and the observed star formation
threshold would have required fine-tuning the parameters of the model.

The fact that equation (\ref{eq:qscaling}) works for both models HSB
and LSB, and 
that $N_H(Q=1)\approx N_H(T=500~\K) \approx N_H(f_{H_2}=10^{-3})$
is consistent with the conclusion that the transition to the cold phase
causes gravitational
instability and that this transition coincides with a sharp increase
in the molecular fraction (i.e., from $f_{H_2} \ll 10^{-3}$ to $f_{H_2} >
10^{-3}$). Using $T=10^3~\K$, or a molecular fraction smaller by one
or two orders of magnitude gives almost identical results. The
hydrogen column densities quoted above can be converted into surface
densities using equation (\ref{eq:sigma-NH}) and into pressures using
equation (\ref{eq:pressure}). For example, the surface density and
pressure corresponding to our fiducial threshold column density, $\log
N_H = 20.75$ (in $\cm^{-2}$), are $\log \Sigma = 0.77$
($\Msun\,\pc^{-2}$) and $\log P/k = 2.6$ ($\cm^{-3}\,\K$). 

Although a critical column density is useful for (semi-)analytic work,
it is of little use if only the volume density is known, as is
generally the case for three-dimensional hydrodynamical simulations.
Fortunately, assuming (approximate) hydrostatic equilibrium, the
critical column density 
can be converted into a critical volume density using equation
(\ref{eq:NJ}), provided that $\mu$ and $T$ are known. For densities
slightly below the threshold value, 
the gas will be warm and nearly fully atomic, and we can
therefore assume $T \approx 10^4~\K$ and $\mu = 1/(1-3Y/4)$ ($\approx
1.22$ for $Y=0.24$). Thus, a physically reasonable star formation
recipe for a hydrodynamical simulation that lacks the physics needed to compute
the transition to the cold phase self-consistently, is to convert gas
elements with $T\sim 10^4~\K$ and density greater than a critical
value computed from equation (\ref{eq:NJ}) and one of the fitting
formulas above. For example, for our fiducial parameters ($f=1$, $Z =
0.1Z_\odot$, $I = 10^6~\cm^{-2}\,\s^{-1}$) the critical density is
$n_{H,{\rm crit}} \approx 4 \times 10^{-2}~\cm^{-3}$.

\section{Comparison with observations}
\label{sec:obs}
In the previous sections it was argued that the transition to the cold
phase leads to gravitational instability and star formation. This
hypothesis leads to several predictions: 
\begin{itemize}
\item Below the threshold surface density both the fraction of gas in the cold
phase and the molecular fraction decline rapidly.
\item The critical surface density depends (weakly) on the gas
fraction, the amount of turbulence, the metallicity, and the intensity
of the UV radiation. For
reasonable parameter values the threshold surface density is $\Sigma_c
\sim 3$--$10~\Msun\,\pc^{-2}$ ($N_{H,{\rm crit}} \sim 3$--$10\times
10^{20}~\cm^{-2}$), which corresponds to a pressure $P/k \sim
10^2$--$10^3~\cm^{-3}\,\K$. 
\item Star formation thresholds are local: SF will occur wherever the
gas surface density exceeds the local threshold value.
\item If the disk has an exponentially declining surface density and
is approximately axisymmetric, then
the disk mass can be predicted if the critical radius $r_c$ and the
disk scale length $R_d$ are measured.
\end{itemize}
In the remainder of this section we will discuss these predictions in
detail and compare them with observations. 

\subsection{Star formation thresholds}
Strong, direct evidence in favor of the hypothesis that the transition
to the cold phase leads to star formation comes
from high-resolution \HI\ observations of 11 nearby spiral galaxies
by Braun (1997). Braun found that the \HI\ emission can be separated
into two distinct components, associated with the cold and
the warm phases respectively: (1) a high-brightness filamentary network
that is marginally resolved at 150~pc and has a velocity FWHM less
than $6~\kms$ and (2) a diffuse interarm and outer disk
component. While the cold component accounts for 60\%--90\% of the
line flux within the star forming disk, its contribution plummets
abruptly near the edge of the optical disk.

Two further predictions of the model are that the critical surface
density should be insensitive to the rotation curve and that the molecular
fraction should rise sharply at the critical surface density. The fact that a
fixed surface density threshold $\Sigma_g \sim 10~\Msun\,\pc^{-2}$
($\log N_H \sim 21$) describes the observations reasonably well, has
been known for some time (e.g., Skillman 1987; Taylor et al.\ 1994;
Ferguson et al.\ 1998b; Leli{\` e}vre \& Roy 2000)
and the results of MK01 confirm this (see their
Fig.~9, \emph{bottom}). This critical value is close to the prediction for our
fiducial parameter values: $\log N_{H,{\rm crit}} \approx 20.75$. It
is interesting to note that the 
observed scatter in $\Sigma_g(r_c)$ is smaller than the scatter
in $Q(r_c)$ (compare Figs.~8 and 9, \emph{bottom}, of MK01). Finally,
MK01's figure 6 demonstrates that the molecular fraction indeed rises
sharply at $\Sigma_g \sim 10~\Msun\,\pc^{-2}$ (see Wong \& Blitz 2002
for additional evidence).

The sample of MK01 contains seven ``subcritical'' disks, i.e.,
galaxies for which $Q(\sigma = 6~\kms)$ never reaches the critical
value in the region of active star formation. However,
inspection of their figure 5 (\emph{bottom}) shows that all but one of
these galaxies have $\Sigma_g \sim 10~\Msun\,\pc^{-2}$ for $r <
r_c$. The one exception 
is NGC\,4698, which is thought to have had its gas removed in a recent
collision with another galaxy (Valluri \& Jog 1990). 
Examples of disks with widespread star formation that are subcritical
according to the K89 criterion, but which have $\Sigma \ga
10~\Msun\,\pc^{-2}$, have also been found in other studies (e.g.,
Thornley \& Wilson 1995; Wong \& Blitz 2002). Thus, observations
indicate that a constant critical surface density works even when the
K89 criterion fails.

Although the critical surface density predicted by the models for the
fiducial parameter values is close to the observed value, the
difference would have been somewhat greater if we had used a radiation
field with an intensity as small as expected for a purely
extragalactic background. In \S\ref{sec:recipes} it was shown that
using the extragalactic value ($\log I \sim 4.5$ photons
$\cm^{-2}\,\s^{-1}$) instead of our fiducial value ($\log I = 6$),
would yield a critical column density that is about a factor of two
lower: $\log N_{H,{\rm crit}} \approx 20.4$.
On the other hand, if we had allowed for some turbulence, the
agreement would have improved. For example, using $f_{th}=0.1$ (note
that the total velocity dispersion is $\sigma \approx
9~\kms\,\sqrt{T_4/f_{th}}$) would increase the critical column density
by about a factor of 2. The bottom line is that because of
uncertainties in the model parameters, as well as systematic errors
due to the limitations of the model, the predictions for the critical
surface density are only accurate to within a factor of a few.

At present it is not even clear whether a threshold surface density of
$\log N_{H,{\rm crit}} \approx 20.4$ is ruled out by the
observations. The reason is that the value of 10~$\Msun\,\pc^{-2}$ is
derived from azimuthally averaged observations. Ferguson et al.\
(1998b; see their Fig.~2) found that although the azimuthally
averaged critical value is of order $10~\Msun\,\pc^{-2}$, individual
\HII\ regions have surface densities as low as 2 to
$4~\Msun\,\pc^{-2}$ ($\log N_H \approx 20.3$ - 20.6).  The reason for
the difference between local thresholds and the thresholds derived
from azimuthally averaged observations could be spiral structure: if
one defines the critical radius as the radius within which SF is
ubiquitous, i.e., not confined to the spiral arms, then the interarm
surface density will be near the threshold value, but the azimuthally
averaged surface density will be larger. 

Note that it does not make sense to measure the critical surface
density without specifying the smoothing scale. After the surface
density increases beyond the threshold value, the gas will fragment
and the surface density of the individual clouds could become very
large. The smoothing scale relevant for
testing global star formation thresholds is probably similar to the
size of \HII\ regions. 

It is interesting to compare the predicted critical column
density for the phase transition with UV absorption line
studies, which find that the transition occurs at $\log N_H
\approx 20.3$--20.8 in the Galaxy (Savage et al.\ 1977) and at $\log
N_H \approx 20.7$--21.2 in the Large Magellanic Cloud (Tumlinson et
al.\ 2002). Using solar metallicity for the Galaxy, equation 
(\ref{eq:fh2scaling}) predicts $\log N_{H,{\rm crit}} \approx 20.5$,
in excellent agreement with the observations. Using a metallicity
$Z=0.2Z_\odot$ for the LMC (e.g., Welty et al.\ 1999), we predict
$\log N_{H,{\rm crit}} \approx 20.7$, which agrees with the lower
range of the observed values. Since the LMC is actively forming stars,
higher values of $f$ and/or $I$ may be more appropriate, in which case
the predicted critical column density would be somewhat higher. 

If gravitational instability is triggered by the drop in the velocity
dispersion associated with the transition to the cold phase, then one
would expect the function $Q(r)$ to be very different depending on
whether the velocity
dispersion is assumed to be constant, as is usually done in observational
studies, or not. One may therefore reasonably ask why the Toomre
criterion, combined with the assumption of a fixed velocity
dispersion, has been reasonably successful in predicting the
critical radius for star formation in spiral galaxies (e.g., K89;
MK01). 

Figure~3 shows a comparison of the real $Q$ parameters (\emph{thick solid
curves}) and those computed under the assumption that $\sigma =
6~\kms$ (\emph{thick dashed curves}). Indeed, it is clear that the
latter assumption 
leads to large errors in the $Q$ parameter. The difference is
particularly large for model LSB, whose entire disk is predicted to be
stable if $\sigma = 6~\kms$. Hence, unless the velocity dispersion is
set to the value appropriate for the transition to the cold phase
($T\sim 10^3~\K$, $\sigma \approx 3~\kms$), assuming a constant
velocity dispersion will yield a critical Toomre parameter different
from unity.

Indeed, MK01, who carried out a thorough
study of star formation thresholds in disk galaxies assuming $\sigma =
6~\kms$, found that the Toomre criterion only works if
one takes $Q\approx 2.0$ as the critical value.  
Figure~3 shows that for both models HSB
and LSB the critical radius, i.e., the radius for which $Q=1$
(\emph{intersection of thick solid curve with solid line}), is almost
identical to 
the radius at which $Q(\sigma = 6~\kms) = 2$ (\emph{intersection of
  thick dashed 
curve with horizontal dashed line}). This remarkable agreement shows that the
model predictions are consistent with the empirical relation of
MK01. Furthermore, because the difference in the assumed
velocity dispersion can account for the claimed difference between the
critical $Q$ values for irregular (HEB98; Hunter et al.\ 2001) and
spiral galaxies (K89; MK01), as discussed in \S\ref{sec:intro}, the  
model predictions also agree with observations of irregular
galaxies.

In short, observations support the hypothesis that the transition to
the cold phase 
determines the critical surface density, and that rotation has little
effect on the star formation threshold in the outer parts of galaxies.

\subsection{Beyond the critical radius}

Since the star formation threshold in the outer disk is insensitive to
the rotation rate, it is essentially a local criterion. Star formation
will only cut off at the same radius throughout the disk if the
surface density and the 
parameters $f$, $Z$, and $I$ are axisymmetric. In particular, beyond
the truncation radius of the stellar disk (i.e., the radius at which
the azimuthally smoothed stellar surface density decreases sharply), star
formation will occur wherever the local gas surface density exceeds
the local threshold value. Observations of
sporadic \HII\ regions beyond the optical radius (e.g., Ferguson et
al.\ 1998b; Brand et al.\ 2001) confirm that star formation thresholds
are a local phenomenon (see also Skillman 1987; Hunter \& Plummer
1996; HEB98; van Zee et al.\ 1997).

LSB galaxies have low star formation efficiencies despite the fact
that their total gas content is normal. In
LSB galaxies the gas is 
spread out over a larger area than in HSB galaxies, and 
consequently the surface density is lower (e.g.,
van der Hulst et al.\ 1993; van Zee et al. 1997). In our model the
star formation threshold is generally not affected by the rotation
rate, and the low rates of star formation in LSB galaxies and in the
far outer parts of HSB galaxies have the same physical cause: the star
formation efficiency is low because the azimuthally averaged gas
surface density is below the threshold value. Although star
formation is suppressed on average, local peaks in the gas surface density
can still give rise to star formation, in agreement with observations
(e.g., van Zee et al.\ 1997; HEB98). We note that this
explanation also holds in the context of other interpretations of the
star formation threshold (e.g., van der Hulst et al.\ 1993; Elmegreen
\& Parravano 1994). 

For many spiral galaxies the truncation of the optical disk is known
to be accompanied by a flaring of the \HI\ disk (e.g., Bottema,
Shostak, \& van der Kruit 1987; Kamphuis \& Briggs 1992).  Our model
can naturally account for the flaring of the gas disk beyond the
truncation radius. The truncation of the stellar disk is caused by the
star formation threshold which, in turn, arises due to the transition
to the cold phase at the critical radius.  The disk thickness is
proportional to the temperature (see eq.\ [\ref{eq:LJ}]), which is
predicted to increase sharply from $T< 10^3~\K$ at $r<r_c$ to $T\sim
10^4~\K$ at $r > r_c$. Hence, naively we would expect the disk
thickness to increase abruptly from $\la 10^2~\pc$ to $\ga 1~\kpc$ at
the edge of the stellar disk. In reality the increase in the disk
scale height will be much smaller and more gradual than the increase
in the temperature because the warm gas just beyond the threshold
feels the full gravity of the neighboring cold gas and stars, and
because the contribution of the dark halo may no longer be negligible
compared to the self-gravity of the disk (i.e., the warm gas has
effectively a low value of $f$). Moreover, within the critical radius
turbulent pressure may increase the disk thickness (i.e., the cold gas
may have a high value of $f$).

The flaring \HI\ layer often exhibits a warp beyond the
stellar disk (e.g., Briggs 1990) and for some galaxies,
notably NGC\,4013, the 
rotation curve exhibits a drop of as much as 10\%--15\% at the edge of the
stellar disk (e.g., Bottema et al.\ 1987; Bottema 1996; van der Kruit
2001). It has been suggested (e.g., Florido et al.\ 2001) that the
small, relatively sudden drop in the rotation curve at the optical
edge of NGC\,4013 poses a problem for theories that try to explain the
observed cut-off in the SFR in terms of a surface
density threshold. However, both observations (Swaters et al.\ 1997;
Schaap, Sancisi, \& Swaters 2000) and numerical models (Struck \&
Smith 1999) indicate that at large scale heights the gas rotates more
slowly. Thus, it seems plausible that the small drop in the rotation
curve of NGC\,4013 is related to the flaring/warping of the disk
(NGC\,4013 has a strong warp). If, as suggested by Swaters et al.\
(1997), the decrease in the circular velocity with scale height is
related to angular momentum conservation, then the effect on the
rotation curve would be minimal if $r^2 \gg L^2$, where $L$ is the
disk thickness. It is therefore interesting that NGC\,4013 has a smaller
truncation radius ($r_c = 9.3~\kpc$; Bottema 1996) than any of the 31
galaxies studied by Pohlen et al.\ (2000a, 2000b).

\subsection{Disk sizes and masses}

Observations indicate that stellar disks end rather abruptly
(e.g., van der Kruit 1979; van der Kruit \& Searle 1981;
Barteldrees \& Dettmar 1994; Fry et
al.\ 1999; Pohlen et al.\ 2000a, 2000b;
de Grijs, Kregel \& Wesson 2001; Florido et al.\ 2001; Kregel, van der
Kruit, \& de Grijs 2002), although the sharpness of the truncation is
still debated (e.g., Pohlen et al.\ 2002). Van der Kruit \&
Searle (1981) measured the  
truncation radius for a sample of 7 edge-on spirals and found that it
occurs at a radius of $4.2 \pm 0.5$ exponential scale lengths ($R_d$) of
the surface brightness distribution. More recently, Pohlen et al.\
(2000a) analyzed a sample of 31 nearby edge-on spiral galaxies and found
$r_c/R_d = 2.9 \pm 0.7$, significantly lower than the value reported
by van der Kruit \& Searle (1981). The Pohlen et al.\ (2000a) sample
is large enough to look for correlations between $r_c$ and other
parameters. Although Pohlen et al.\ (2000a) did
not find a correlation between $r_c/R_d$ and Hubble type, they did
find an anti-correlation with $R_d$: large disks have relatively
shorter cut-off radii. This anti-correlation can probably account for
the differences between the results of van der Kruit \& Searle (1981)
and Pohlen et al.\ (2000a), as the sample of the latter authors
contains more galaxies with large scale lengths. 

If the critical radius for star formation remains fixed over a
substantial part of the star formation history of the galaxy, then the star
formation threshold will create a cut-off in the stellar surface
density. The optical edge will in that case roughly coincide with the
critical radius, provided that the galaxy formed inside-out. 

From equation (\ref{eq:expdensprof}) it can be seen that for the MMW98
disk model the critical radius is given by 
\begin{equation}
{r_c \over R_d} = \ln {m_d M_{200} \over 2\pi \Sigma_c} - 2 \ln R_d.
\label{eq:size}
\end{equation}
The critical radius depends on the critical surface density
$\Sigma_c$, the disk mass $m_d M_{200}$, and the disk scale length
$R_d$. Note that there is a degeneracy in the models between
increases in the critical surface density and decreases in the disk
mass. If the ratio $m_d M_{200}/\Sigma_c$ is fixed, then $r_c/R_d$ is
anti-correlated with $R_d$.

\begin{figure}
\begin{center}
\epsscale{1.1}
\plotone{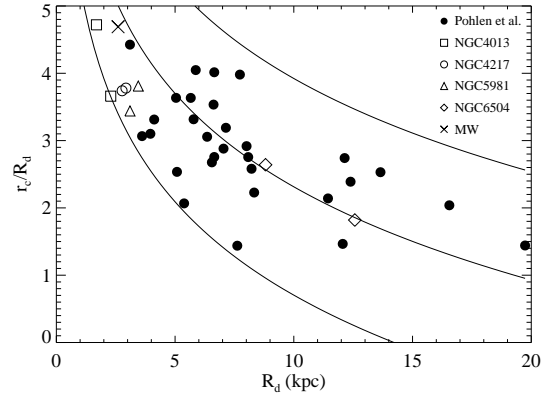}
\figcaption[f5.eps]{Disk size relative to disk scale length as a
function of disk scale length, showing the sample of edge-on galaxies
of Pohlen et al.\ (2000b; \emph{solid points}), data taken from 
Florido et al.\ (2001; \emph{open symbols}; there is one data point
for either side of the 
galaxy), data for the Galaxy from Freudenreich (1998; \emph{cross}),
and the model predictions for $\log N_{H,{\rm crit}} =
20.75$ (\emph{solid curves}) and $M_d/7.5\times 10^9~\Msun = 1$
(\emph{top}), 5 (\emph{middle}), and
25 (\emph{bottom}). The models can clearly fit the observations for a
reasonable range of disk masses.}
\end{center}
\end{figure}

Figure~5 (\emph{filled circles}) shows the observations of Pohlen et al.\
(2000b). The anti-correlation between $r_c/R_d$ and $R_d$ reported by
Pohlen et al.\ (2000a) is clearly visible\footnote{Although the
observed anti-correlation between $r_c/R_d$ and $R_d$ 
is suggestive, it should be kept in mind that it could be the
result of measurement errors in $R_d$. Since disks are generally only well
described by purely exponential profiles over at most a few scale
lengths, and since the 
truncation radius is often not well defined, the measured values of
$R_d$ will depend somewhat on the definition of $r_c$ (e.g., Pohlen
2001).}. The solid curves are the 
predictions of our model for our fiducial value of $\log N_{H,{\rm
crit}} = 20.75$ ($\log \Sigma_c \approx 0.77 ~\Msun\,\pc^{-2}$) and,
from top to bottom, disk masses of 1, 5, and 25 times $7.5\times
10^9~\Msun$, respectively. The models can clearly reproduce
the observed scaling of the size of the optical disk with the
exponential scale length.

Figure~5 (\emph{open symbols}) shows the results of Florido et al.\
(2001), who observed four edge-on spiral galaxies in the near
infrared. Florido et al.\ measured the truncation radius and the
disk scale length separately for each side of the galaxies, and hence
there are two data points per galaxy in figure~5. While the two sides
of the galaxies NGC\,4217 and NGC\,5981 give almost 
identical results, there
are significant differences between the two sides for NGC\,4013 and
NGC\,6504. What is interesting is that even though the two sides fall in
different parts of the $r_c/R_d$-$R_d$ diagram, they can both be fitted
by a model with the same disk mass.

If the critical radius $r_c$ and the disk scale length $R_d$ have both
been measured, then equation (\ref{eq:size}) can be used to predict
the corresponding 
disk mass, which is proportional to the assumed critical surface
density $\Sigma_c$. Thus, if an independent estimate of the disk mass
is available, then this provides an additional test of the model. For
the Milky Way robust estimates of all three quantities have been
published. Figure~5 ({\emph cross}) shows the measurements of Freudenreich
(1998): $R_d \approx 2.6~\kpc$ and $r_c/R_d \approx 4.7$. 
Equation (\ref{eq:size}) then yields a disk mass $M_d \approx 2.7\times
10^{10}~\Msun$, remarkably close to the results of Dehnen \&
Binney (1998), who found $M_d = 3$--$5\times 10^{10}~\Msun$ from
various observational constraints.

A consistency check is possible even if no measurements of the disk
mass are available: we can compute the 
mass-to-light ratio of the disk and compare it with expectations from
population synthesis models. The total mass-to-light
ratio of the disk in the $X$ band, $\Upsilon_X$, is given by
\begin{equation}
2.5\log\Upsilon_X = M_X - M_X(\odot) + 2.5\log M_d/M_\odot,
\end{equation}
where $M_X$ and $M_X(\odot)$ are the absolute magnitudes of the galaxy
and the sun respectively. Population synthesis models yield \emph{stellar}
mass-to-light ratios, whereas the above equation predicts the
\emph{total} mass-to-light ratio.
While the surface density of the disk is generally dominated by stars
in the inner 
parts of the galaxy, their contribution is very small for $r>r_c$. Since the 
fraction of the total disk mass exterior to $r_c$ is not always
negligible for an exponential disk, we need to 
use the disk mass interior to $r_c$, $M_d(r_c) = [1 - \exp(-r_c/R_d) -
r_c/R_d\exp(-r_c/R_d)]M_d$, when computing $\Upsilon$. This still
gives only an upper limit to the stellar mass-to-light 
ratio, because it includes the mass in gas.

Figure~6 shows the resulting mass-to-light ratio in the $B$ band for
the sample of Pohlen et al.\ (2000b), computed using the distances
listed by those authors, the apparent 
magnitudes from the NASA Extragalactic Database\footnote{
http://nedwww.ipac.caltech.edu/} (corrected for Galactic extinction
using the results of Schlegel, Finkbeiner, \& Davis 1998), and
$M_B(\odot) = 5.48$. The
mass-to-light ratios scatter between 1 and 5.5, with a mean of
3.1. Note that all points can be moved up and down in proportion to
the critical surface density (we used our fiducial value $\log
N_{H,{\rm crit}} = 20.75$). The scatter may not be real, since there
are (unknown) errors in the measurements of
$r_c$, $R_d$, $m_B$, the distances, and of course in the
model. Furthermore, there may be small differences in the real critical
surface densities from one galaxy to the next, depending on the
appropriate values of $f$, $Z$, and $I$. Nevertheless, the predicted
values for the total mass-to-light ratio are in accord with the
predictions of population synthesis models for the stellar
mass-to-light ratio in the $B$ band (e.g., Bell \& de Jong
2001). Hence, we conclude that the model predicts disk masses that
are consistent with the observations.

\subsection{Uncertainties}
\label{sec:uncertainties}

Although the agreement between the model and observations is
encouraging, it is important to keep in mind that there are
considerable uncertainties in both the observational results and the
theoretical predictions.  

The
observables required to test the Toomre criterion are the rotation
curve, the surface
densities in atomic and molecular gas, and the velocity
dispersion. The latter two quantities are most difficult to measure
with sufficient accuracy. 

One important source of error for all observables is
azimuthal smoothing. Azimuthal smoothing would be harmless if
galaxies were perfectly axisymmetric, but the presence of spiral
structure proves that they are not. Although the models are
axisymmetric, the star formation threshold criterion does
not require this. The fate of a perturbation, which does not have to
be ringlike, depends on the local rotation rate and the local
surface density. Hence, for non-axisymmetric galaxies the critical
radius is not well defined, and individual regions beyond the
azimuthally smoothed threshold can form stars if their surface
density exceeds the local threshold value.

The molecular gas surface densities are derived from CO maps, using a
conversion factor that has been calibrated using local, Galactic
observations, i.e., using gas with a metallicity that may be
considerably higher than in the outer disk. The CO maps rarely cover
the outer parts of the disk and the \Htwo\ column near the optical
radius is therefore generally estimated by extrapolating the trend
measured at smaller radii. If the models are correct, then the
molecular fraction rises sharply shortwards of the critical radius
(see Fig.~2), and extrapolating the \Htwo\ surface density would
result in an overestimate of the threshold surface density.

\begin{figure}
\begin{center}
\epsscale{1.1}
\plotone{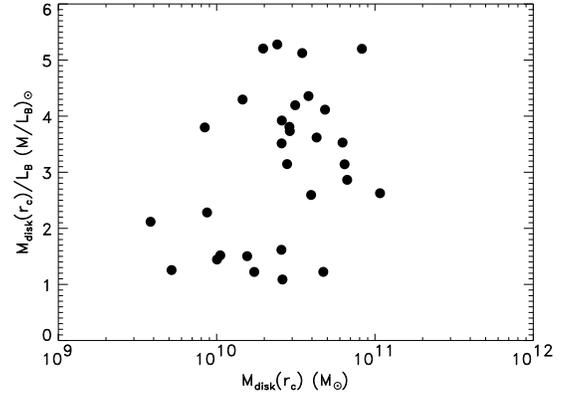}
\figcaption[f6.eps]{The predicted mass-to-light ratio in
the $B$-band versus the predicted disk mass interior to the critical
radius for the sample of Pohlen et al.\ (2000b). These mass-to-light
ratios are based on the total disk mass 
and are thus upper limits for the stellar mass-to-light
ratios. Disk masses are computed from the measured critical radii and
disk scale lengths, and are proportional to the assumed critical
surface density. The results fall within the range expected for
$B$-band stellar mass-to-light ratios (e.g., Bell \& de Jong 2001),
indicating that the disk masses predicted by the model are consistent
with the observations.} 
\end{center}
\end{figure}

The velocity dispersion is very difficult to measure and is 
therefore usually assumed to be constant. 
To test whether, as argued here, the star
formation threshold is set by the transition to the cold phase, one
would like to measure the change in the velocity dispersion. However,
even if one could measure the dispersion, the interpretation would not
be straightforward because the ISM will be multiphase shortwards of the
critical radius, whereas it is only the velocity dispersion of the cold
phase that is relevant. More importantly, it is likely that there is no sharp
decrease in the velocity dispersion because feedback from star
formation will generate turbulence, which could lead to
self-regulation (i.e., $Q\approx 1$). Note that this is fully
consistent with the hypothesis that the transition to the cold phase
is responsible for the observed cut-off in the star formation
rate. All that this hypothesis requires, is that the velocity
dispersion is small in the absence of star formation.

Of course, the model also suffers from uncertainties. Although the
assumptions of plane-parallel radiative transfer, a step-function
vertical density profile, thermal equilibrium, and hydrostatic
equilibrium do capture most of the relevant physics, they are
approximations. Note that the latter assumption does not even hold
exactly in our model: even though we compute physical quantities like
the density, temperature and disk thickness as a function of radius,
we ignore the effects of gradients in these quantities when doing
so. Nevertheless, it is demonstrated in the appendix that the model
accurately predicts the disk thickness of an isothermal, exponential
disk. On the other hand, the rapid change in the temperature at the
critical radius probably causes us to overestimate the sharpness of
the decrease in the scale height (the thin, cold disk decreases the
effective $f$-value for the neighboring warm, thick disk). Moreover,
for the warm, outer disk the effect of the dark halo may no longer be
negligible compared to the self-gravity of the disk (again, this would
decrease the $f$-value for the warm disk). However, 
the phase transition should remain sharp, because the gas is thermally
unstable at intermediate temperatures, and because self-shielding
and the formation of molecular hydrogen provide positive feedback
loops. 

Our model does not include several physical processes that could in
principle trigger star formation in the otherwise warm, outer
disk. Examples are swing-amplifier instabilities and spiral density
waves. If these processes are important, then they could 
undermine the conclusion that the phase transition triggers
gravitational instability. However, to form a cold, molecular phase,
regions of enhanced density will still have to exceed the critical
surface density and the phase transition will still trigger
instability on small scales. Although gravity-driven turbulence
seems unavoidable, 
we note that recent simulations (e.g, Wada et al.\ 2002) indicate that
in the extended disk, 
beyond the critical radius for star formation, the typical turbulent
velocity dispersion 
is substantially below the typical observed value ($8~\kms$), which,
in our model, corresponds to the thermal velocity dispersion of the
warm phase.

Perhaps even more important than the uncertainties in the model
assumptions are the uncertainties in the model parameters, in
particular $f$, $Z$, and $I$. In \S\ref{sec:pars} we motivated our
fiducial values ($f=1$, $Z=0.1~Z_\odot$, $I = 10^6~{\rm
photons}~\cm^{-2}\,\s^{-1}$). In \S\ref{sec:recipes} it was shown that
the critical surface density is fairly insensitive to the values of
these parameters. Typically, a parameter has to change by a factor
$\sim 10^3$ for the threshold surface density to change by a factor
10. Nevertheless, the large uncertainty in the appropriate values for
the turbulent pressure and the intensity of the UV radiation imply
that the predictions could be off by a factor of a few.

\section{Conclusions}
\label{sec:conclusions}

Observations indicate that disk galaxies have rather sharp edges: both
the SFR and the stellar surface density decline
sharply beyond a few disk scale lengths. These observations are
usually explained in terms of a star formation threshold, set by the
Toomre criterion for gravitational instability in a thin, rotating
disk (e.g., Spitzer 1968; K89). Neither rotation nor pressure can
stabilize the disk if the 
Toomre $Q$ parameter ($Q\equiv c_s\kappa/\pi G\Sigma_g$) is smaller
than unity or, equivalently, if the surface density exceeds the
critical value $c_s\kappa/\pi G$. Since it is
difficult to measure the velocity dispersion $\sigma$ (and thus the
sound speed $c_s$), observational studies usually assume a fixed
value. Provided that the threshold $Q$ value is allowed to differ from
unity, and to be different for spirals and irregulars, the Toomre
criterion appears reasonably successful. However, observations of
``subcritical disks'' (i.e., $Q>Q_c$ everywhere) with widespread star
formation and of \HII\ regions beyond the critical radius [$=
r(Q=Q_c)$] have cast doubt on the general applicability of a $Q$
threshold. As an alternative, Elmegreen \& Parravano (1994) have
emphasized that the need for a cold phase to form stars introduces a
minimum critical pressure, which depends on the radiation field and
the metallicity. They argued that high $Q$ values prevent star
formation indirectly because they inhibit the formation of regions
with sufficiently high pressures to contain a multiphase ISM.

To investigate the physical cause of the observed star formation
threshold in the outer parts of galaxies, a model was constructed for
a gaseous, exponential disk 
embedded in a dark halo. The disk is self-gravitating, contains metals
and dust, and is illuminated by UV radiation. It was found that the
critical surface density for the transition to the cold phase agrees
with empirically derived constant star formation thresholds (e.g.,
Skillman 1987), which therefore supports the idea of Elmegreen \&
Parravano (1994) that a cold phase is critical to star formation (see
also HEB98, Hunter et al.\ 2001, Billett et al.\ 2002, and Elmegreen
2002).  

In the models the drop in the thermal velocity dispersion associated with the 
transition from the warm ($T\sim 10^4~\K$) to the cold ($T< 10^3~\K$)
phase causes the disk to become gravitationally unstable on a large
range of scales, which
suggests that this phase transition plays a more important role than
had hitherto been recognized. Since stars form in molecular clouds, it
is likely that the presence of a cold, gravitationally unstable phase
will lead to star formation. 
The transition to the cold phase is sharp because the gas is thermally
unstable at intermediate temperatures and because both self-shielding
and the formation of molecular hydrogen provide positive feedback
loops. The phase transition is associated with a sharp
increase in the molecular fraction (from $f_{H_2} \ll 10^{-3}$ to
$f_{H_2} > 10^{-3}$), in agreement with observations. 

Galactic rotation, which includes both the Coriolis force and shear,
does not affect the critical surface density at which the phase
transition occurs and cannot stabilize the cold phase in the outer
disk.  It is, however, important to note that insofar as the Toomre
$Q$ parameter controls the formation of global perturbations such as
spiral arms and bars, it can have an indirect effect on the ability of
the disk to form regions in which the surface density exceeds the
threshold value for the transition to the cold phase (Elemgreen \&
Parravano 1994), which could then trigger gravitational instability on
smaller scales.

While turbulence ultimately has a stabilizing effect, it can also
promote local collapse by generating density enhancements.  
Turbulence driven by feedback from star formation should not
be taken into account when computing a global star formation
threshold. However, there exists several physical mechanisms that
could drive turbulence even in the absence of star formation, but
which were not included in the models. Examples are infalling
gas clouds, shocks from spiral density waves, and the magnetorotational
instability. We argued that it is unlikely that turbulence dominates
the velocity dispersion of the gas in the extended \HI\ disks, which
can be accounted for entirely by thermal motions provided it is kept
warm by the (extragalactic) UV radiation, as predicted by the model.

Although the presence of strong turbulent support would increase the
critical surface density, it would not
undermine the conclusion that the transition to the cold phase triggers
instability. The reason is that turbulent support increases the
surface density required for the phase transition by about as much as
it increases the velocity dispersion of the gas, thereby leaving the
$Q$ parameter at the phase transition nearly unchanged. In the models
the phase transition triggers instability for turbulent pressures as
large as 250 times the thermal pressure. Such a high turbulent
pressure would give rise to velocity dispersions that are ruled out by
measurements of \HI\ line widths in the outer parts of galaxies.

Although the model used here can predict the threshold surface
density above which some of the gas will become cold and unstable to
fragmentation, it cannot predict which fragments will succeed in
forming stars. Moreover, once stars form, feedback processes will
create a complex, multiphase ISM that is not well described by the simple
model. Hence, the predicted critical surface densities should be
interpreted as global star formation thresholds. 

The critical radius for star formation can be defined as the radius
within which the azimuthally averaged surface density exceeds the
critical value required for the transition to the cold phase
(e.g., Elmegreen \& Parravano 1994). However,
since rotation is generally unimportant, star
formation thresholds are a local phenomenon: peaks in the
surface density that exceed the threshold value will form stars
regardless of their position in the disk. Thus, sporadic star
formation beyond $r_c$ is possible, in particular in spiral
arms (e.g., van Zee et al.\ 1997; HEB98). Compared to high surface 
brightness galaxies, low surface brightness galaxies have low star
formation efficiencies because their (azimuthally averaged) surface
density becomes subcritical at smaller radii relative to their disk
scale length (e.g., van der Hulst et al.\ 1993). 

The model can
account for the radii and surface densities at which the SFR
is observed to cut off. Model predictions regarding the
scaling of the size of the stellar disk with the disk mass and
scale length agree with the observations. In particular, the
mass-to-light ratios predicted on the basis of measured values of the
critical radius, the disk scale length, and the luminosity, are
consistent with expectations from population synthesis models.

Scaling relations were computed for the dependence of the critical
column density on the intensity of the UV-radiation, the metallicity,
the relative 
importance of thermal and turbulent pressure, and the mass fraction in
stars and dark matter. Assuming hydrostatic 
equilibrium, the critical column density can be converted into an
equivalent critical pressure or volume density. These scaling
relations can be used to 
implement star formation thresholds in (semi-)analytic models and
three-dimensional hydrodynamical simulations of galaxy
formation. For reasonable
parameter values, we find a critical surface density $\Sigma_c \sim
3$--$10~\Msun\,\pc^{-2}$ ($N_{H,{\rm crit}} \sim 3$--$10\times
10^{20}~\cm^{-2}$), which 
corresponds to a pressure $P/k \sim 10^2$--$10^3~\cm^{-3}\,\K$, and a
volume density $n_{H}\sim 10^{-2}$--$10^{-1}~\cm^{-3}$.

Thus, the hypothesis that the transition to the cold phase triggers
gravitational instability and star formation, leads to a large number
of predictions that appear to be supported by available
observations. It provides an explanation for a range of observed
phenomena and correlations, some of which were thought to be
unrelated. Future observations, as well as numerical simulations that
include radiative transfer, molecules, dust, and metals, could test
these predictions in more detail and help refine the model.

\acknowledgments 
It is a pleasure to thank Anthony Aguirre, John Bahcall, Crystal
Martin, David Weinberg, and the anonymous referee for a 
careful reading of the manuscript. This work was supported by grants
from the W.~M.~Keck foundation and the National Science Foundation
(PHY-0070928). 

\appendix

In this section it is demonstrated that for the case of a gaseous
isothermal exponential disk, equation (\ref{eq:NJ}) is a very good 
approximation to the exact solution. The density profile for an
isothermal exponential disk is, 
\begin{equation}
\rho(r,z) = {\Sigma(r) \over 2H(r)} \mbox{sech}^2\left ({z \over
H(r)}\right ), 
\end{equation}
where the scale height $H$ is given by
\begin{equation}
H = {c_s^2 \over \pi G \Sigma}
\end{equation}
(Spitzer 1942). For the purpose of the radiative transfer calculation, the 
characteristic density at radius $r$ is the column density
weighted density 
\begin{equation} 
\left <\rho \right >_N \equiv {1 \over \Sigma} \int \rho^2 dz
= {\Sigma \over 3H}.
\end{equation}
Ignoring the small mass contained in metals, the total hydrogen column
density $N_H = \Sigma (1-Y)/m_H$ can then be written as,
\begin{equation} 
N_H = \left ({3 \over \pi}\right )^{1/2} \left ({\gamma k \over \mu
m_H^2 G}\right )^{1/2} (1-Y)^{1/2} n_H^{1/2} T^{1/2},
\end{equation}
where $n_H \equiv \left <\rho\right >_N (1-Y)/m_H$ is the
characteristic hydrogen number density. This  
expression differs from equation (A5) of Schaye (2001a; our eq.\
[\ref{eq:NJ}]), which was
derived by equating the sound crossing time and the dynamical time, by
a factor $\sqrt{3/\pi} \approx 0.98$.


\begin{thebibliography}{}
\bibitem[Barteldrees \& Dettmar(1994)]{1994A&AS..103..475B} 
Barteldrees, A.~\& Dettmar, R.-J.\ 1994, \aaps, 103, 475

\bibitem[Bell \& de Jong(2001)]{2001ApJ...550..212B} 
Bell, E.~F.~\& de Jong, R.~S.\ 2001, \apj, 550, 212

\bibitem[Billett, Hunter, \& Elmegreen(2002)]{2002AJ....123.1454B} Billett, 
O.~H., Hunter, D.~A., \& Elmegreen, B.~G.\ 2002, \aj, 123, 1454 

\bibitem[Binney \& Tremaine(1987)]{1987gady.book.....B} 
Binney, J., \& Tremaine, S. 1987, Galactic Dynamics (Princeton:
Princeton Univ.\ Press)

\bibitem[Black(1987)]{1987ip...symp..731B} 
Black, J.~H.\ 1987, ASSL Vol.~134: Interstellar Processes, 731

\bibitem[Bland-Hawthorn \& Maloney(1999)]{1999hvc..work..212B} 
Bland-Hawthorn, J.~\& Maloney, P.~R.\ 1999, ASP Conf.~Ser.~166:
Stromlo Workshop on High-Velocity Clouds, 212

\bibitem[Bland-Hawthorn \& Maloney(2002)]{}
Bland-Hawthorn, J.~\& Maloney, P.~R.\ 2002, ASP Conf.~Ser.~254:
Extragalactic Gas at Low Redshift, 267

\bibitem[Borriello \& Salucci(2001)]{2001MNRAS.323..285B} Borriello, A.~\&
Salucci, P.\ 2001, \mnras, 323, 285

\bibitem[Bottema(1996)]{1996A&A...306..345B} 
Bottema, R.\ 1996, \aap, 306, 345

\bibitem[Bottema, Shostak, \& van der Kruit(1987)]{1987Natur.328..401B} 
Bottema, R., Shostak, G.~S., \& van der Kruit, P.~C.\ 1987, \nat, 328, 401

\bibitem[Brand et al.(2001)]{2001A&A...377..644B} 
Brand, J., Wouterloot, J.~G.~A., Rudolph, A.~L., \& de Geus, E.~J.\
2001, \aap, 377, 644

\bibitem[Braun(1997)]{1997ApJ...484..637B} 
Braun, R.\ 1997, \apj, 484, 637

\bibitem[Briggs(1990)]{1990ApJ...352...15B} 
Briggs, F.~H.\ 1990, \apj, 352, 15

\bibitem[Corbelli \& Salpeter(1995)]{1995ApJ...450...32C} 
Corbelli, E.~\& Salpeter, E.~E.\ 1995, \apj, 450, 32

\bibitem[Corbelli, Galli, \& Palla(1997)]{1997ApJ...487L..53C} 
Corbelli, E., Galli, D., \& Palla, F.\ 1997, \apjl, 487, L53

\bibitem[de Grijs, Kregel, \& Wesson(2001)]{2001MNRAS.324.1074D} 
de Grijs, R., Kregel, M., \& Wesson, K.~H.\ 2001, \mnras, 324, 1074

\bibitem[Dehnen \& Binney(1998)]{1998MNRAS.294..429D} 
Dehnen, W.~\& Binney, J.\ 1998, \mnras, 294, 429

\bibitem[Elmegreen(1993)]{} 
Elmegreen, B.~G.\ 1993, in Star-Forming Galaxies and Their
Interstellar Media, ed.\ J. Franco, F. Ferrini, \& G. 
Tenorio-Tagle (Cambridge: Cambridge Univ.\ Press), 337

\bibitem[Elmegreen(2002)]{2002ApJ...577..206E} Elmegreen, B.~G.\ 2002, 
\apj, 577, 206 

\bibitem[Elmegreen \& Parravano(1994)]{1994ApJ...435L.121E} 
Elmegreen, B.~G.~\& Parravano, A.\ 1994, \apjl, 435, L121

\bibitem[Fall \& Efstathiou(1980)]{1980MNRAS.193..189F} 
Fall, S.~M.~\& Efstathiou, G.\ 1980, \mnras, 193, 189

\bibitem[Ferguson, Gallagher, \& Wyse(1998)]{1998AJ....116..673F} 
Ferguson, A.~M.~N., Gallagher, J.~S., \& Wyse, R.~F.~G.\ 1998a, \aj, 116, 673

\bibitem[Ferguson et al.(1998)]{1998ApJ...506L..19F} 
Ferguson, A.~M.~N., Wyse, R.~F.~G., Gallagher, J.~S., \& Hunter,
D.~A.\ 1998b, \apjl, 506, L19 

\bibitem[Ferland(2000)]{2000RMxAC...9..153F} 
Ferland, G.~J.\ 2000, Revista Mexicana de Astronomia y Astrofisica
Conference Series, 9, 153

\bibitem[Ferland et al.(1998)]{1998PASP..110..761F} 
Ferland, G.~J., Korista, K.~T., Verner, D.~A., Ferguson, J.~W.,
Kingdon, J.~B., \& Verner, E.~M.\ 1998, \pasp, 110, 761

\bibitem[Field(1965)]{1965ApJ...142..531F} Field, G.~B.\ 1965, \apj,
142, 531

\bibitem[Florido et al.(2001)]{2001A&A...378...82F} 
Florido, E., Battaner, E., Guijarro, A., Garz{\' o}n, F., \& Jim{\'
e}nez-Vicente, J.\ 2001, \aap, 378, 82

\bibitem[Freudenreich(1998)]{1998ApJ...492..495F} 
Freudenreich, H.~T.\ 1998, \apj, 492, 495

\bibitem[Fry et al.(1999)]{1999AJ....118.1209F} 
Fry, A.~M., Morrison, H.~L., Harding, P., \& Boroson, T.~A.\ 1999,
\aj, 118, 1209

\bibitem[Gerritsen \& Icke(1997)]{1997A&A...325..972G} 
Gerritsen, J.~P.~E.~\& Icke, V.\ 1997, \aap, 325, 972

\bibitem[Goldreich \& Lynden-Bell(1965)]{1965MNRAS.130..125G} 
Goldreich, P.~\& Lynden-Bell, D.\ 1965, \mnras, 130, 125

\bibitem[Haardt \& Madau(2001)]{}
Haardt, F., \& Madau, P. 2001, in 21st Moriond Astrophys.\ Meeting,
Clusters of Galaxies and the High-Redshift Universe Observed in
X-rays: Recent Results of \emph{XMM-Newton} and \emph{Chandra}, ed.\
D. M. Neumann \& J. T. T. Van (Paris: Editions Frontieres)

\bibitem[Henry \& Worthey(1999)]{1999PASP..111..919H}
Henry, R.~B.~C.~\& Worthey, G.\ 1999, \pasp, 111, 919

\bibitem[Hunter \& Plummer(1996)]{1996ApJ...462..732H} Hunter, D.~A.~\& 
Plummer, J.~D.\ 1996, \apj, 462, 732 

\bibitem[Hunter, Elmegreen, \& Baker(1998)]{1998ApJ...493..595H}
Hunter, D.~A., Elmegreen, B.~G., \& Baker, A.~L.\ 1998, \apj, 493, 595
(HEB98)

\bibitem[Hunter, Elmegreen, \& van Woerden(2001)]{2001ApJ...556..773H} 
Hunter, D.~A., Elmegreen, B.~G., \& van Woerden, H.\ 2001, \apj, 556, 773

\bibitem[Kamphuis \& Briggs(1992)]{1992A&A...253..335K} 
Kamphuis, J.~\& Briggs, F.\ 1992, \aap, 253, 335

\bibitem[Kennicutt(1989)]{1989ApJ...344..685K} 
Kennicutt, R.~C.\ 1989, \apj, 344, 685 (K89)

\bibitem[Kregel, van der Kruit, \& de Grijs(2002)]{2002MNRAS.334..646K} 
Kregel, M., van der Kruit, P.~C., \& de Grijs, R.\ 2002, \mnras, 334, 646 

\bibitem[Leli{\` e}vre \& Roy(2000)]{2000AJ....120.1306L} 
Leli{\` e}vre, M., \& Roy, J. 2000, \aj, 120, 1306 

\bibitem[Lo, Sargent, \& Young(1993)]{1993AJ....106..507L} Lo, K.~Y., 
Sargent, W.~L.~W., \& Young, K.\ 1993, \aj, 106, 507 

\bibitem[Maloney(1993)]{1993ApJ...414...41M} 
Maloney, P.\ 1993, \apj, 414, 41

\bibitem[Martin \& Kennicutt(2001)]{2001ApJ...555..301M} 
Martin, C.~L.~\& Kennicutt, R.~C.\ 2001, \apj, 555, 301 (MK01)

\bibitem[Meurer, Carignan, Beaulieu, \& Freeman(1996)]{1996AJ....111.1551M} 
Meurer, G.~R., Carignan, C., Beaulieu, S.~F., \& Freeman, K.~C.\ 1996, \aj, 
111, 1551 

\bibitem[Mo, Mao, \& White(1998)]{1998MNRAS.295..319M} 
Mo, H.~J., Mao, S., \& White, S.~D.~M.\ 1998, \mnras, 295, 319 (MMW98)

\bibitem[Navarro, Frenk, \& White(1997)]{1997ApJ...490..493N} 
Navarro, J.~F., Frenk, C.~S., \& White, S.~D.~M.\ 1997, \apj, 490, 493
(NFW)

\bibitem[Oh \& Haiman(2002)]{2002ApJ...569..558O} Oh, S.~P.~\& Haiman, 
Z.\ 2002, \apj, 569, 558

\bibitem[Olling(1995)]{1995AJ....110..591O} 
Olling, R.~P.\ 1995, \aj, 110, 591

\bibitem[Pettini et al.(1997)]{1997ApJ...478..536P} 
Pettini, M., King, D.~L., Smith, L.~J., \& Hunstead, R.~W.\ 1997,
\apj, 478, 536 

\bibitem[Pettini et al.(1999)]{1999ApJ...510..576P} 
Pettini, M., Ellison, S.~L., Steidel, C.~C., \& Bowen, D.~V.\ 1999,
\apj, 510, 576

\bibitem[Pohlen(2001)]{}
Pohlen, M.\ 2001, PhD Thesis, Ruhr-University Bochum, Germany

\bibitem[Pohlen, Dettmar, \& L{\" u}tticke(2000a)]{2000A&A...357L...1P} 
Pohlen, M., Dettmar, R.-J., \& L{\" u}tticke, R.\ 2000a, \aap, 357, L1

\bibitem[Pohlen et al.(2000b)]{2000A&AS..144..405P} 
Pohlen, M., Dettmar, R.-J., L{\" u}tticke, R., \& Schwarzkopf, U.\
2000b, \aaps, 144, 405

\bibitem[Pohlen et al.(2002)]{}
Pohlen, M., Dettmar, R.-J., L{\" u}tticke, R., \& Aronica, G.\
2002, \aap, 392, 807

\bibitem[Quirk(1972)]{1972ApJ...176L...9Q} 
Quirk, W.~J.\ 1972, \apjl, 176, L9

\bibitem[Romeo(1992)]{1992MNRAS.256..307R} 
Romeo, A.~B.\ 1992, \mnras, 256, 307

\bibitem[Safronov(1960)]{1960AnAp...23..979S} 
Safronov, V.~S.\ 1960, Annales d'Astrophysique, 23, 979

\bibitem[Savage et al.(1977)]{1977ApJ...216..291S} 
Savage, B.~D., Drake, J.~F., Budich, W., \& Bohlin, R.~C.\ 1977, \apj,
216, 291

\bibitem[Schaap, Sancisi, \& Swaters(2000)]{2000A&A...356L..49S} 
Schaap, W.~E., Sancisi, R., \& Swaters, R.~A.\ 2000, \aap, 356, L49

\bibitem[Schaye(2001a)]{2001ApJ...559..507S} 
Schaye, J.\ 2001a, \apj, 559, 507

\bibitem[Schaye(2001b)]{2001ApJ...562L..95S} 
Schaye, J.\ 2001b, \apjl, 562, L95

\bibitem[Schlegel, Finkbeiner, \& Davis(1998)]{1998ApJ...500..525S} 
Schlegel, D.~J., Finkbeiner, D.~P., \& Davis, M.\ 1998, \apj, 500, 525

\bibitem[Scott et al.(2002)]{} 
Scott, J., Bechtold, J., Morita, M., Dobrzycki, A., \& Kulkarni,
V. 2002, ApJ, 571, 665

\bibitem[Sellwood \& Balbus(1999)]{1999ApJ...511..660S} 
Sellwood, J.~A.~\& Balbus, S.~A.\ 1999, \apj, 511, 660

\bibitem[Skillman(1987)]{1987sfig.conf..263S} 
Skillman, E.~D.\ 1987, in Star Formation in Galaxies, edited by
C. J. Londsdale Persson (NASA Conf.\ Pub.\ CP-2466), p.\ 263

\bibitem[Spitzer(1942)]{1942ApJ....95..329S} 
Spitzer, L.~J.\ 1942, \apj, 95, 329

\bibitem[Spitzer(1968))]{1968dms..book.....S} 
Spitzer, L. 1968, Diffuse Matter in Space (New York: Wiley) 

\bibitem[Struck \& Smith(1999)]{1999ApJ...527..673S} 
Struck, C.~\& Smith, D.~C.\ 1999, \apj, 527, 673

\bibitem[Swaters, Sancisi, \& van der Hulst(1997)]{1997ApJ...491..140S} 
Swaters, R.~A., Sancisi, R., \& van der Hulst, J.~M.\ 1997, \apj, 491, 140

\bibitem[Taylor et al.(1994)]{1994AJ....107..971T} 
Taylor, C.~L., Brinks, E., Pogge, R.~W., \& Skillman, E.~D.\ 1994,
\aj, 107, 971

\bibitem[Thornley \& Wilson(1995)]{1995ApJ...447..616T} 
Thornley, M.~D.~\& Wilson, C.~D.\ 1995, \apj, 447, 616

\bibitem[Toomre(1964)]{1964ApJ...139.1217T} 
Toomre, A.\ 1964, \apj, 139, 1217

\bibitem[Tumlinson et al.(2002)]{2002ApJ...566..857T} 
Tumlinson, J.~et al.\ 2002, \apj, 566, 857

\bibitem[Valluri \& Jog(1990)]{1990ApJ...357..367V} 
Valluri, M.~\& Jog, C.~J.\ 1990, \apj, 357, 367

\bibitem[van den Bosch \& Swaters(2001)]{2001MNRAS.325.1017V} 
van den Bosch, F.~C.~\& Swaters, R.~A.\ 2001, \mnras, 325, 1017

\bibitem[van der Hulst et al.(1993)]{1993AJ....106..548V} 
van der Hulst, J.~M., Skillman, E.~D., Smith, T.~R., Bothun, G.~D.,
McGaugh, S.~S., \& de Blok, W.~J.~G.\ 1993, \aj, 106, 548

\bibitem[van der Kruit(1979)]{1979A&AS...38...15V} 
van der Kruit, P.~C.\ 1979, \aaps, 38, 15

\bibitem[van der Kruit(1988)]{1988A&A...192..117V} 
van der Kruit, P.~C.\ 1988, \aap, 192, 117

\bibitem[van der Kruit(2001)]{2001gddg.conf..119V} 
van der Kruit, P.~C.\ 2001, ASP Conf.~Ser.~230: Galaxy Disks and Disk
Galaxies, 119 

\bibitem[van der Kruit \& Searle(1981)]{1981A&A....95..105V} 
van der Kruit, P.~C.~\& Searle, L.\ 1981, \aap, 95, 105

\bibitem[van Zee et al.(1997)]{1997AJ....113.1618V}
van Zee, L., Haynes, M.~P., Salzer, J.~J., \& Broeils, A.~H.\ 1997,
\aj, 113, 1618

\bibitem[Vladilo(1998)]{1998ApJ...493..583V} 
Vladilo, G.\ 1998, \apj, 493, 583

\bibitem[Wada, Meurer, \& Norman(2002)]{2002ApJ...577..197W} Wada, K., 
Meurer, G., \& Norman, C.~A.\ 2002, \apj, 577, 197 

\bibitem[Welty et al.(1999)]{1999ApJ...512..636W} 
Welty, D.~E., Frisch, P.~C., Sonneborn, G., \& York, D.~G.\ 1999,
\apj, 512, 636

\bibitem[Wong \& Blitz(2002)]{2002ApJ...569..157W} Wong, T.~\& Blitz, L.\ 
2002, \apj, 569, 157

\end{thebibliography}
\end{document}